\documentclass[pra,showpacs,twocolumn]{revtex4}
\usepackage{amsmath,amssymb}
\usepackage{enumerate}
\usepackage{color}
\usepackage{cancel}

\begin{document}

\title{Generalization of the Fermi Pseudopotential}

\author{Trang T. L\^e}
\affiliation{Department of Mathematics, University of Tulsa, Tulsa, Oklahoma 
74104, USA}
\author{Zach Osman}
\affiliation{Department of Mathematics, University of Tulsa, Tulsa, Oklahoma 7
4104, USA}
\author{D. K. Watson}
\affiliation{Department of Physics and Astronomy, University of Oklahoma, 
Norman, Oklahoma 73019, USA}
\author{Martin Dunn}
\affiliation{Asp St., Norman, Oklahoma 73019, USA}
\author{B. A. McKinney}
\affiliation{Tandy School of Computer Science, University of Tulsa, Tulsa, 
Oklahoma 74104, USA}
\affiliation{Department of Mathematics, University of Tulsa, Tulsa, Oklahoma 
74014, USA}
\date{\today}


\begin{abstract}
Introduced eighty years ago, 
the Fermi pseudopotential has been a powerful concept 
in multiple fields of physics.  
It replaces the detailed shape of a 
potential by a delta-function operator multiplied by a parameter giving the 
strength of the potential. For Cartesian dimensions $d>1$, 
a regularization operator is necessary 
to remove singularities in the wave function. In this study, we develop a Fermi 
pseudopotential generalized to $d$ dimensions (including non-integer) and to 
non-zero wavenumber, $k$. Our approach has the advantage of circumventing 
singularities that occur in the wave function at certain integer values of 
$d$ while being valid arbitrarily close to integer $d$. In the limit of 
integer dimension, we show that our generalized pseudopotential is equivalent 
to previously derived $s$-wave pseudopotentials. Our pseudopotential 
 generalizes the operator to non-integer 
dimension, includes energy ($k$) dependence, and
simplifies the dimension-dependent coupling constant expression derived from 
a Green's function approach.  We apply this pseudopotential 
to the problem of two cold atoms ($k\to0$) in a harmonic trap and extend the 
energy expression to arbitrary dimension. 
\end{abstract}
\maketitle

\section{Introduction}

The Fermi pseudopotential has had a profound influence since its introduction
in 1936\cite{fermi}, having been used 
extensively in
nuclear, condensed matter, atomic and chemical physics. Pseudopotentials
 replace the details of
short-range interaction potentials with shape-independent potentials that 
reproduce the physics often through a single parameter. 
When the range of
interaction is small compared to the interparticle distance, the details of the
interaction become unimportant allowing the use of pseudopotentials.
 Early use of the Fermi pseudopotential in the context of nuclear 
physics\cite{breit1,breit2,bethe,blatt}, 
recognized that in three dimensions, a regularizing operator must be combined
with the delta function to remove singularities in the wave function. 
The shape independence of the
Fermi pseudopotential
has been embodied in an effective range expansion\cite{bethe}
that is the basis of 
far-reaching applications in multiple fields of physics. Applications of the
Fermi pseudopotential to
manybody systems was initiated
by Huang and Yang and applied to systems of 
hardcore bosons\cite{huang1,huang2,huang3,wu}. 

The Fermi pseudopotential, also referred to as a  zero-range or contact 
potential,  has seen a resurgence of use in the current
literature of ultra-cold atom gases. The 
 experimental realization of BEC led to the study of ultracold
systems of atoms with long de Broglie wavelengths that permitted the use 
of  simple
pseudopotentials since the details of the interaction potential were not
relevant. These pseudopotentials for ultracold gases are typically
characterized by a single parameter, the scattering length.

Although the great majority of work using pseudopotentials has been for
three dimensional systems, the Fermi pseudopotential has also been used in
other integer dimensions, including one, two
and five dimensions. The one-dimensional Fermi 
pseudopotential, i.e. the delta function, has been used extensively as an
exactly solvable model problem of manybody theory\cite{horiuchi}. It 
has also been applied to quark tunneling\cite{cleric}, H photodetachment 
and multiphoton 
ionization\cite{berson,lagattuta,sanpera}, 
 periodic lattices\cite{grossmann1}, in the theory of 
transport phenomena\cite{avishai} and for two cold atoms in a harmonic
trap\cite{busch}. 
The five dimensional Fermi pseudopotential
has been used in the problem of narrow resonances in infinite,  
uniform chains\cite{grossmann2} and as well as for two cold atoms in a 
harmonic trap\cite{busch}. 

Two dimensions is of particular interest since
several interesting collective phenomena can emerge in reduced dimensionality 
including the Berezinsky-Kosterlitz-Thouless 
transition\cite{berezinskii,kosterlitz,hadzibabic},
quantum magnetism\cite{sachdev}, high-temperature 
superconductivity\cite{wen} and  topological insulators\cite{hasan}. 
Two dimensional systems have been suggested as
potential 
platforms for topological quantum computation\cite{nayak}.

More recently the unprecedented control of ultracold systems has resulted in
highly anisotropic trapping configurations yielding
quasi-one and quasi-two dimensional systems\cite{Ketterle01, bloch}. When these ultracold
particles are tightly confined along two directions and weakly
 confined along the third, a quasi one-dimensional system is created. 
The fractional quantum Hall
effect in quasi-two dimensions may be possible for bosons confined tightly 
along one direction and 
weakly along the other two. These quasi one- and two- dimensional configurations
can be combined with a tunable scattering length to lead to largely
unexplored physical regimes. These configurations have been studied by
renormalizing the coupling strength\cite{bolda}, and by explicitly
developing a smooth pseudopotential in 2D\cite{whitehead}, but an interesting
alternate 
approach would be to
use a Fermi pseudopotential for  non-integer dimensions between one and two, 
perhaps specified by the ratio of the different coupling constants.

The Fermi pseudopotential has previously 
been extended to all the odd integers\cite{wod}.
 General
expressions for all the even integers $d > 2$ have not been explicitly given 
due to a complicated singularity structure which results in expressions
that are not generally useful\cite{wod}.
In this paper, we present a generalization of the
Fermi pseudopotential to all values of the dimension, both even and
odd integers, as well as non-integer values which are obtained as 
a natural result of our approach.

 Our method has the advantage of circumventing 
singularities that occur at certain integer values of $d$,
simplifying the dimension-dependent coupling constant expression derived from 
an earlier Green's function approach\cite{wod}, generalizing
 the pseudopotential to non-integer 
dimension and including energy ($k$) dependence.
This work thus offers a broader approach to obtaining valid pseudopotentials in
any dimension of interest.  

This paper is organized in the following way. In Section \ref{sec:one} we 
discuss the singularities that occur at integer values of $d$ and our
strategy to circumvent these singularities by staying at non-integer values
of dimension and using non-integer or 
``fractional'' derivatives to develop a general pseudopotential.
 Appendix  \ref{sec:frac_calc} reviews the 
relevant mathematical 
properties of fractional calculus that are used in our formulation of 
the generalized pseudopotential operator.

In Section \ref{sec:3_frac_pseudo}, we construct the generalized 
pseudopotential for
two cases:  $k \to 0$ and finite $k$.
In Appendix \ref{small_k}, we construct the general $d$ pseudopotential
for $k \to 0$.  This development closely parallels the $3d$ development found
in Huang's Statistical Mechanics textbook\cite{huang}. With the help of
 Appendices \ref{A_SchrodingerSolDeriv} and \ref{appendixB}, 
we construct the general-dimension pseudopotential 
with non-zero wavenumber $k$ (Eq. \ref{pseudoPotential}). We derive this 
general pseudopotential by introducing an arbitrary dimension Frobenius 
series solution to the $d$-dimensional relative-motion Schr\"{o}dinger 
free-wave equation with a hard-sphere boundary condition.
 In Section \ref{sec:4_pseudo_limits} 
and Appendix \ref{appendixC}, we show that integer limits of the general 
pseudopotential, with zero and non-zero $k$, agree with previous 
generalizations. Our pseudopotential simplifies the expression for the 
dimension-dependent coupling constant in an earlier Green's function 
derivation~\cite{wod}, and includes 
energy ($k$) dependence. In Section \ref{sec:5_two_cold} along with 
Appendices \ref{appendixD} and \ref{appendixE}, we apply our general 
pseudopotential to the problem of two cold atoms in a harmonic 
trap~\cite{busch} and generalize the Busch energy relationships to 
arbitrary $d$ (Eqs. \ref{eq:energyEquation} 
and \ref{eq:approxEnergyEquation}). Section \ref{sec:discuss} contains 
discussion and possible future directions.

\section{Pseudopotentials and singularities}\label{sec:one}

Pseudopotentials are shape-independent (contact) potentials that approximate 
the effect of detailed short-range interaction potentials.
 The simplest pseudopotential is the $\delta$-function, introduced 
by Fermi~\cite{fermi}. Later, it was noted that, for three dimensions a 
regularizing operator of the form $(\frac{\partial}{\partial r})r$ is required 
to remove singularities in the wave function\cite{breit1,breit2,bethe,blatt}. 
For $d>3$, not only must the leading singularity of the form $1/r^{d-2}$ be 
removed by the pseudopotential, but it must also remove other lower-order 
singularities in $1/r$. For example, when $d=7$, 
singularities $1/r^5$, $1/r^3$, and $1/r$ arise and must be removed.
When $d$ is an even integer, $\ln(r)$ singularities arise in the series 
solution resulting in expressions that become quite complicated as even $d$
values increase.  
The Green's function approach used by W\'odkiewicz~\cite{wod} to 
derive the arbitrary odd integer-$d$ pseudopotential removes all singularities 
in higher dimensions but does not include energy dependence. In more strongly 
interacting regimes, energy dependence becomes important, and it has been 
shown, for example, that using an energy-dependent scattering length improves 
the description of two-atom systems~\cite{blume,bolda}. Our generalized 
pseudopotential removes singularities in higher dimensions and includes energy 
dependence. 

In the current study, we consider non-integer dimensions and demonstrate 
agreement with previous results in the limit of integer dimensionality.
Our strategy to circumvent this singularity structure is to
work only with non-integer dimensions. This approach bypasses these 
singularities because $d$ is never allowed to be integer, although
it may be arbitrarily 
close. Developing the formalism for non-integer dimension pseudopotentials
requires the use of non-integer calculus, specifically the ability to take
derivatives at non-integer order. This branch of mathematics has developed
 several alternative definitions for  non-integer'' or
``fractional'' derivatives.  In Appendix \ref{sec:frac_calc} we review the
properties of the fractional derivative that we have chosen based on  the
physics to develop our pseudopotentials. After deriving our pseudopotential, 
we demonstrate 
agreement with previous results in the limit of integer dimensionality.


\section{Construction of the generalized pseudopotential}\label{sec:3_frac_pseudo}
The purpose of the regularized Fermi pseudopotential is to obtain a Schr\"{o}dinger equation with an inhomogeneous term that reproduces the effect of the hard-sphere boundary condition on the scattered wave function. That is, we wish to derive a pseudopotential that places a node in the wave function at $r=\bar{a}$, where $\bar{a}$ is small compared to the interatomic spacing. For $r>\bar{a}$, the system follows a non-interacting $d$-dimensional relative-motion Schr\"{o}dinger equation for two atoms interacting via a hard sphere of radius $\bar{a}$:
\begin{equation}
\label{Schrodinger}
\left(\nabla_d^2+k^2\right)\psi\left(r\right)=0, \quad r>\bar{a}
\end{equation}
where $\nabla_d^2 = \frac{1}{r^{d-1}}\frac{d}{dr}r^{d-1}\frac{d}{dr}$. The quantity $\bar{a}$ is a radius in $d$ dimensions. For notational simplicity, we write $\bar{a}$ instead of $\bar{a}_d$ when the dimension is clear from the context. When the dimension is not obvious, we use a subscript ({\it e.g.}, $\bar{a}_3$). In the following sections, we derive the pseudopotential for $k\to0$ from the boundary condition of the Schr\"{o}dinger equation (Section \ref{subsec:smallk}) and for arbitrary $k$ by solving the Schr\"{o}dinger equation (Eq. \ref{Schrodinger}) for $\psi(r)$ (Section~\ref{sec:3.2}).

\subsection{General $d$ pseudopotential with $k \to 0$ from the boundary condition}\label{subsec:smallk}

We derive the $k \to 0$ general-$d$ (non-integer) pseudopotential 
in some detail because many of the techniques used are similar to the more 
complicated arbitrary-$k$ solution. Similar to Ref.~\cite{huang} 
(Huang pp. 276-277 for $d=3$), we integrate the $d$-dimensional 
Schr\"odinger equation (Eq.~\ref{Schrodinger}) when $k \to 0$ and the 
wave function satisfies the boundary condition:
\begin{equation}
\label{boundaryCondition}
\psi(r) \to \left(1-\frac{\bar{a}^{d-2}}{r^{d-2}}\right)B \quad \textrm{as} \,\,
 r \to 0,
\end{equation}
where $B$ is an integration constant that depends on the boundary condition 
as $r\to \infty$, $d$ is the spatial dimensionality and $k$ is the magnitude 
of the relative wave vector defined by $E=\hbar^2k^2/2\mu$, with $E$ being the relative-motion energy and $\mu=m/2$  the reduced mass of the system.

To isolate $B$ in terms of $\psi$ as $r\to0$, one can operate on Eq. (\ref{boundaryCondition}) with $r^{d-2}$ and then a derivative of order $d-2$.
In Appendix \ref{small_k} using our choice for non-integer derivatives, 
we obtain the following expression for the
generalized pseudopotential at  $k \to 0$:

\begin{equation}
\label{pseudoSmallk}
_{C_0}V_{k\to0}^{(d)} (r) = \frac{\Omega(d) \bar{a}^{d-2}}{\Gamma(d-2)}\delta^{(d)}(r){\; _{C_0}D}_r^{d-2}r^{d-2}.
\end{equation}
Eq. (\ref{pseudoSmallk}) will be the limiting case ($k\to0$) of the more 
general $k$-dependent pseudopotential (Eq. \ref{pseudoPotential}) in the 
next section, and it will be the starting point for the ultra-cold 
two-trapped-atom derivation in Section \ref{sec:5_two_cold}.

\subsection{General $d$ pseudopotential for arbitrary energy $k$}\label{sec:3.2}
When $k$ is non-zero, one cannot simply integrate the Schr\"odinger equation. To derive the pseudopotential we first develop the power series solution of the Schr\"odinger equation to identify the singularities that need to be removed. Solving the differential equation of Eq. (\ref{Schrodinger}) and developing a series solution for $\psi$ using a non-integer $d$ Frobenius method (see Eqs.~\ref{eq:A1}--\ref{eq:A9} for details), we find a regular solution
\begin{equation}\label{regularSolForm}\psi_{reg}(r)=\sum^{\infty}_{n=0}\alpha_nr^{2n}\end{equation}
and an irregular solution
\begin{equation}\label{irregularSolForm}
\psi_{irreg}(r)=c\ln(r)\psi_{reg}+\frac{1}{r^{d-2}}\sum^\infty_{n=0}\beta_nr^{2n}.
\end{equation}
In general, the coefficient $c$ depends on the dimension $d$ (see Appendix \ref{A_SchrodingerSolDeriv} for all cases). However, since we assume non-integer $d$, we can allow $c=0$ and avoid the $\ln(r)$ term (see discussion near Eqs.~\ref{eq:A11} and \ref{eq:betan_appendixA}). The non-integer-$d$ series solution can be made arbitrarily close to integer, and we will find that the limit for odd $d$ is smooth while the even $d$ limit must be handled more carefully. 

Thus, for non-integer dimension, we may then combine Eqs. (\ref{regularSolForm}) and (\ref{irregularSolForm}) to obtain the 
non-integer $d$ solution to the Schr\"odinger equation:
\begin{equation}\label{generalSolForm}
\psi(r)=\sum_{n=0}^\infty\alpha_{2n}r^{2n}+\frac{1}{r^{d-2}}\sum_{n=0}^\infty\beta_{2n}r^{2n}.
\end{equation}
Similar to the approach for $k \to 0$ 
(Section \ref{subsec:smallk} and Appendix \ref{small_k}), we integrate the left side of the Schr\"{o}dinger equation Eq. (\ref{Schrodinger}) over a small sphere of infinitesimal radius $\epsilon$ about the origin. However, now $k$ is allowed to be nonzero and we use $\psi$ in Eq. (\ref{generalSolForm}).  We find that as $\epsilon \to 0$ (Appendix \ref{appendixB})
%
\begin{equation}
\int\left(\nabla_d^2+k^2\right)\psi\left(r\right) dV= \Omega(d)\beta_0(2-d)\int \delta^{(d)}(r)dV
\end{equation}
or (from Eq.~\ref{pseudoBeta0}) 
\begin{equation}\label{eq:int_SE_beta0}
\left(\nabla^2_d + k^2\right)\psi = \Omega(d) \beta_{0}(2-d) \delta^{(d)}(r).
\end{equation}

Now our goal is to find an explicit expression for $\beta_0$ from the series solution. We do this by choosing an operator that will give $\beta_0$ in terms of the full wave function as $r \to 0$.  We first operate on Eq. (\ref{generalSolForm}) with $r^{d-2}$ and a derivative of order $d-2$ to remove singularities when $r\to0$:
\begin{equation}
\begin{split}
&\bigl[_{C_0}D_r^{d-2}r^{d-2}\psi\bigr]_{r\to0} \\
&= \Bigl[_{C_0}D_r^{d-2}\bigl(\sum_{n=0}^\infty\alpha_{2n}r^{2n+d-2} 
+\sum_{n=0}^\infty\beta_{2n}r^{2n}\bigr)\Bigr]_{r\to0} \\
&= \Bigl[\sum_{n=0}^\infty
	\frac{\Gamma(2n+d-1)}{\Gamma(2n+1)}\alpha_{2n}r^{2n} \\
&+ \sum_{n=\lfloor\frac{d}{2}\rfloor}^\infty
	\frac{\Gamma(2n+1)}{\Gamma(2n-d+3)}\beta_{2n}r^{2n-d+2}\Bigr]_{r\to0},
\end{split}
\end{equation}
where in the second line we used a fractional derivative equation for powers 
(See Eq. \ref{RLpower}). When $r \to 0$, we note the summation on the right does not contribute because of the lower limit $\lfloor\frac{d}{2}\rfloor$ (floor of $d/2$) and only the $n=0$ term contributes from the left summation so that
\begin{equation}\label{eq:D_alpha0}
\left[_{C_0}D_r^{d-2}r^{d-2}\psi\right]_{r\to0}  =  \Gamma(d-1) \alpha_0. 
\end{equation}
We wish to rewrite $\beta_0$ in terms of the wave function $\psi$ to substitute into Eq. (\ref{eq:int_SE_beta0}). Recall the relationship between $\alpha_0$ and $\beta_0$ (Eq. \ref{alpha0beta0}) derived from imposing the boundary condition (node at $r = \bar{a}$) for the Schr\"odinger equation:
\begin{equation}
\alpha_0= - 
\frac{\beta_0 }{\bar{a}^{d-2}}
\frac
	{_0F_1\left(;2-\frac{d}{2}; - \frac{k^2\bar{a}^2}{4}\right)}
	{ {_0F}_1\left(;\frac{d}{2}; - \frac{k^2\bar{a}^2}{4}\right)},
\end{equation}
where $_0F_1(;\cdot; \cdot)$ is the confluent hypergeometric function~\cite{handbookMathFuncs}. Using this expression for $\alpha_0$ to write Eq. (\ref{eq:D_alpha0}) in terms of $\beta_0$, we find
\begin{equation}
\left[_{C_0}D_r^{d-2}r^{d-2}\psi\right]_{r\to0}  =  - \Gamma(d-1) 
\frac{\beta_0 }{\bar{a}^{d-2}}
\frac
	{_0F_1\left(;2-\frac{d}{2}; - \frac{k^2\bar{a}^2}{4}\right)}
	{ {_0F}_1\left(;\frac{d}{2}; - \frac{k^2\bar{a}^2}{4}\right)},
\end{equation}
and we can isolate $\beta_0$:
\begin{equation}
\label{mybeta0}
\beta_0 = - 
\frac{\bar{a}^{d-2}}{\Gamma(d-1) } 
\frac
	{_0F_1\left(;\frac{d}{2}; - \frac{k^2\bar{a}^2}{4}\right)}
	{ {_0F}_1\left(;2-\frac{d}{2}; - \frac{k^2\bar{a}^2}{4}\right)}
\left[_{C_0}D_r^{d-2}r^{d-2}\psi\right]_{r\to0}.
\end{equation}
Finally, substituting this $\beta_0$ in Eq. (\ref{eq:int_SE_beta0}), we obtain the finite $d$, finite $k$ pseudopotential
\begin{equation}
\label{prePseudoPotential}
\begin{split}
&\left(\nabla_d^2+k^2\right)\psi\left(r\right)\\
& = \frac{\Omega(d)(d-2)\bar{a}^{d-2}}{\Gamma(d-1)} \, T(d,k\bar{a}) \, \delta^{(d)}(r)\,{\; _{C_0}D}_r^{d-2}r^{d-2}\psi,
\end{split}
\end{equation}
where we define 
\begin{equation}
\label{TrangEquation}
T(d,k\bar{a}) = \frac{_0F_1\left(;\frac{d}{2}; - \frac{k^2\bar{a}^2}{4}\right)}{_0F_1\left(;2-\frac{d}{2}; - \frac{k^2\bar{a}^2}{4}\right)},
\end{equation}
which is related to the scattering phase shift in $d$-dimensions for $s$-wave. The case $T(3,k\bar{a})=\tan(k\bar{a})/k\bar{a}$ leads to the correct $s$-wave scattering phase shift for a hard sphere with radius $\bar{a}$ in $d=3$. Higher dimensions have more complicated forms, which we discuss along with other limits below. We note that $T(d,k\bar{a})=0$ for even $d>2$ ($T=1$ for $d=2$) and requires a perturbation expansion of Eq. (\ref{TrangEquation}) near even $d>2$.

Simplifying the operator on the right side of Eq. (\ref{prePseudoPotential}), we obtain the generalized pseudopotential in non-integer dimension,
\begin{equation}
\label{pseudoPotential}
_{C_0}V_{k}^{(d)} = \frac{\Omega(d)\bar{a}^{d-2}}{\Gamma(d-2)} \, T(d,k\bar{a}) \, \delta^{(d)}(r)\,{\; _{C_0}D}_r^{d-2}r^{d-2}.
\end{equation}
In the following sections, we explore cases of the generalized pseudopotential, test its validity in various limits of $k$ and $d$, and apply it to the problem of two ultra-cold atoms in a harmonic trap in arbitrary dimension.

\section{Cases of the generalized pseudopotential for specific limits of $d$ and $k$}\label{sec:4_pseudo_limits}
Here we demonstrate some specific cases of the $s$-wave, arbitrary-$d$ pseudopotential (Eq. \ref{pseudoPotential}). We show the analytical forms of the pseudopotential for $d=3$ and $d=5$ with zero and non-zero $k$.  We also derive the relationship between the simple coupling constant in our generalized pseudopotential with the more complicated expression in Ref.~\cite{wod}.

\subsection{Non-zero $k$ with $d =3$ and $d=5$}
When the non-integer dimension approaches $d \to 3$, the finite $k$ generalized pseudopotential (Eq. \ref{pseudoPotential}) takes the form
\begin{equation}
_{C_0}V_{k}^{(3)} = \frac{\Omega(3)\bar{a}}{\Gamma(1)} \, T(3,k\bar{a}) \, \delta^{(3)}(r)\,{\; _{C_0}D}_{r}r, 
\end{equation}
where the phase-shift related function (Eq. \ref{TrangEquation}) becomes
\begin{equation}
T(3,k\bar{a})=\frac{\tan(k\bar{a})}{k\bar{a}}
\end{equation}
and yields
\begin{equation}
\label{pseudoPotential3}
_{C_0}V_{k}^{(3)} = \frac{4\pi}{k\cot(k\bar{a})}\delta^{(3)}(r)\frac{d}{d r}r,
\end{equation}
which is consistent with the $s$-wave pseudopotential derived by Huang (Eq. (13.12) in Ref.~\cite{huang}). For $d\to3$, the prefactor that comes from $T$ is positive and increases with $k\bar{a}$ and diverges as $k\bar{a} \to \pi/2$. Different behavior emerges for $d\to5$ next. 

Similarly, when the non-integer dimension approaches $d \to 5$, the finite $k$ generalized pseudopotential (Eq. \ref{pseudoPotential}) takes the form
\begin{equation}
_{C_0}V_{k}^{(5)} = \frac{\Omega(5)\bar{a}^{3}}{\Gamma(3)} \, T(5,k\bar{a}) \, \delta^{(5)}(r)\,{\; _{C_0}D}_r^{3}r^{3},
\end{equation}
where the phase-shift related function (Eq. \ref{TrangEquation}) is
\begin{equation}
T(5,k\bar{a})=\frac{3}{(k\bar{a})^3} \left( \frac{\tan(k\bar{a})-k\bar{a}}{k\bar{a}\tan(k\bar{a})-1} \right)
\end{equation}
and results in
\begin{equation}
\label{pseudoPotential5}
_{C_0}V_{k}^{(5)} = \frac{4\pi^2}{k^3} \left( \frac{\tan(k\bar{a})-k\bar{a}}{k\bar{a}\tan(k\bar{a})-1} \right) \delta^{(5)}(r) \frac{d^3}{d r^3}r^3.
\end{equation}
As $k\bar{a}$ increases, the expression in parentheses for $d\to5$ above is a negative and decreasing function of $k\bar{a}$ and diverges as $k\bar{a} \to 0.86033$ (first positive numerical solution). 

\subsection{The $k \to 0$ limit of the arbitrary $d$ pseudopotential}
For $d \to 3$, the series expansion about $k\to0$ of the phase-shift related function (Eq. \ref{TrangEquation}) gives the usual result for the hard-sphere potential:
\begin{equation}
T(3,k\bar{a})=\frac{\tan(k\bar{a})}{k\bar{a}} \approx 1 + \frac{k^2\bar{a}^2}{3} + O(k^4).
\end{equation}
Thus, in the $k \to 0$ (cold-atom) $d\to3$ limit, the generalized pseudopotential,
\begin{equation}
\label{pseudoSmallkd3}
_{C_0}V_{k\to0}^{(3)}=4\pi \bar{a} \delta^{(3)}(r)\frac{d}{dr}r,
\end{equation}
agrees with Huang (Eq. (13.11) in Ref.~\cite{huang}).

For non-even $d$, the series expansion about $k\to0$ of the phase-shift related function (Eq. \ref{TrangEquation}) is
\begin{equation}
T(d,k\bar{a}) \approx 1 - \frac{(d-2)}{(d-4)}\frac{k^2\bar{a}^2}{d} + O(k^4)
\end{equation}
and the generalized pseudopotential (Eq. \ref{pseudoPotential}) approaches 
\begin{equation}
\label{eq:generalPseudo_k0}
_{C_0}V_{k\to0}^{(d)} = \frac{\Omega(d)\bar{a}^{d-2}}{\Gamma(d-2)}\delta^{(d)}(r){\; _{C_0}D}_r^{d-2}r^{d-2}
\end{equation}
which is the $k \to 0$ pseudopotential (Eq. \ref{pseudoSmallk}) previously derived from the boundary condition of the Schr\"{o}dinger equation. 

\subsection{Relationship between the $k \to 0$ generalized pseudopotential $_{C_0}V_{k\to0}^{(d)}$ and the W\'odkiewicz pseudopotential (Green's function derivation) $V_{W}^{(d)}$ }
The operator in Eq.~(\ref{eq:generalPseudo_k0}) has the same form as the operators derived by W\'odkiewicz for odd $d$ (Eqs. 5.4-5.5 in Ref.~\cite{wod}). However, it is not obvious how the prefactors are related because of the complicated summation (Eq. 5.5 Ref.~\cite{wod}) and the way the coupling constant $a_d$ is defined in Ref.~\cite{wod}. The coupling $a_d$ is not simply the hard sphere radius in $d$ dimensions like our $\bar{a}_d$. For each dimension, $a_d$ is related to the $1d$ scattering length $a_1$. Here we show how W\'odkiewicz's $a_d$ is related to $\bar{a}_d$ and how the W\'odkiewicz prefactors (Eq. 5.5 Ref.~\cite{wod}) can be simplified. 

First, in Appendix \ref{AppendixC1} we show that the complicated regularizing coefficient $\gamma_{2n+1}$ (Eq. 5.5 in Ref.~\cite{wod}) 
\begin{equation}
\gamma_{2n+1}=\frac{\pi^{-1/2}\Gamma\left(\frac{1}{2}-n\right)}{\sum_{l=0}^{n-1}(-1)^{l+n}2^{-l+n}\frac{(n-1+l)!(2n-1)!}{l!(n-1-l)!(n+l)!}}
\end{equation}
simplifies to
\begin{equation}
\gamma_{2n+1}=\frac{1}{\Gamma(2n)}=\frac{1}{\Gamma(d-1)}.
\end{equation}
Next, we show how W\'odkiewicz's $a_d$ is related to $\bar{a}_d$. The coupling constant or area of the potential $a_{d}$ in Ref.~\cite{wod} is defined in terms of the scattering amplitude (Eq. 3.1 in Ref.~\cite{wod}):
\begin{equation}f_k^{(d)}=f_k^{(2n+1)}=-\frac{i^{-n} k^{n-1} (2 n-1)\text{!!} }{k^{2 n-1}+i  2^{n-1} \frac{\pi ^n (2 n-1)\text{!!}\hbar}{  a_{2 n+1} \alpha}}.
\end{equation} 
The scattering amplitude equation in Ref.~\cite{wod} contains a typographical error, which we note in Appendix \ref{typoAppendix}. The area of the potential $a_{d}$ is determined by finding the pole of the scattering amplitude $f^{(d)}_k$ in the physical $k$ half plane while keeping the energy at each single bound state in $d$-dimensions constant. In Ref.~\cite{wod}, expressions for $a_d$ are only derived for $d=1,3,$ and $5$, but here we extend to all odd $d$ the relationship between $a_d$ and the $1$-dimensional area of the potential (see Eq.~\ref{eq:appendC_W_ad}):
\begin{equation}
a_{d}=\frac{(-2 \pi )^{(d-1)/2} (d-2)\text{!!} \hbar^{2d-2}}{a_1^{d-2} m^{d-1}}, \quad \quad d \; \textrm{odd}.
\end{equation} 
Substituting our general formula above for the strength or ``area of potential,'' $a_d$, into W\'odkiewicz's definition of the pseudopotential in terms of $a_d$ (Eq. (1.2) in Ref.~\cite{wod}),
\begin{equation}
V^{(d)}_{W}(r) = -a_d\delta^{(d)}(r)\hat R_d, 
\end{equation} 
where $\hat R_d$ is the regularizing operator (Eq. (5.4) in Ref.~\cite{wod}), we produce an explicit general expression for the W\'odkiewicz Fermi pseudopotential for odd $d>1$ (Eq.~\ref{eq:appendC_VW_converted}):
\begin{equation}
\label{WodPseudo}
\begin{split}
V^{(d)}_{W}(r)& = -\frac{\hbar^{2d-2}(-2 \pi )^{(d-1)/2} (d-4)\text{!!} }{m^{d-1}a_1^{d-2} } \\
&\quad \times \delta ^{(d)}(r) \frac{1}{\Gamma (d-2) }\frac{\partial^{d-2}}{\partial r^{d-2}} r^{d-2}, \quad d>1 \; \textrm{odd},
\end{split}
\end{equation} 
where we remind the reader that $(-1)\text{!!}=1$.
\begin{table}
\begin{center}
	\begin{tabular}{p{1cm}|p{4.5cm}|p{3.0cm}}
		 & $V^{(d)}_{W}(\vec{r})$  & ${_{C_0}V}_{k\to0}^{(d)} (r)$  \\[0.5em]
		 \hline\hline
		 &\\[-1em]

		$d\to1$ & $-a_1 \delta^{(1)}(r)$ & $-\frac{2}{\bar{a}_1} \delta ^{(1)}(r)$ \\[0.8em] \hline
				 &\\[-1em]


		$d\to3$ & $\displaystyle\frac{2\pi\hbar^4}{m^2a_1} \delta ^{(3)}(r) \frac{\partial}{\partial r} r$ & $\displaystyle4\pi \bar{a}_3 \delta ^{(3)}(r) \frac{\partial}{\partial r} r$ \\[0.8em] \hline
				 &\\[-1em]
		$d\to5$ & $\displaystyle-\frac{8\hbar^8\pi^{5/2}}{3m^4a_1^3\Gamma\left(-\frac{3}{2}\right)} \delta ^{(5)}(r) \frac{\partial^{3}}{\partial r^{3}} r^{3}$& $\displaystyle\frac{4\pi^2}{3}\bar{a}_5^3 \delta ^{(5)}(r) \frac{\partial^{3}}{\partial r^{3}} r^{3}$ \\[0.9em] \hline
	\end{tabular}
\end{center}
	\caption{Comparison for dimensions $1, 3$ and $5$ of the W\'odkiewicz pseudopotential $V^{(d)}_{W}(\vec{r})$ (Eq.~\ref{WodPseudo}) and the $k\to0$ limit of the general $d$ pseudopotential ${_{C_0}V}_{k\to0}^{(d)} (r)$ (Eq.~\ref{eq:generalPseudo_k0} from the limit of Eq.~\ref{pseudoPotential}). The pseudopotentials are written in terms of W\'odkiewicz $1d$ scattering length, $a_1$, and our $d$-dimensional wave function node radius, $\bar{a}_d$. The two columns are equivalent when the conversion (Eq.~\ref{eq:abar_a1_conversion}) between $a_1$ and $\bar{a}_d$ is used. }
	\label{comparison1}
\end{table}

In the approach in \cite{wod}, $a_1$ is the 1-dimensional area of potential, and the pseudopotential coupling constant in other dimensions is related to $a_1$. This causes $a_1$ to appear in the denominator of Eq. (\ref{WodPseudo}) for $d>1$. In contrast, we use the characteristic length $\bar{a}_d$ of the potential, which causes the repulsive strength of ${_{C_0}V}_{k\to0}^{(d)} (r)$ (Eq. \ref{eq:generalPseudo_k0} from Eq. \ref{pseudoPotential}) to be directly proportional to $\bar{a}_d$ for $d>1$. We verified (Table~\ref{comparison1}) that the limit of ${_{C_0}V}_{k\to0}^{(d)} (r)$ when $d$ approaches an odd integer agrees with the pseudopotential results listed in Ref.~\cite{wod} ($d=1,3,$ and $5$). Equating Eqs. (\ref{eq:generalPseudo_k0}) and (\ref{WodPseudo}) we find that our wave function node position $\bar{a}_d$ and the constant $a_1$ used by W\'odkiewicz~\cite{wod} are related by the following expression
\begin{equation}\label{eq:abar_a1_conversion}
\bar{a}_d^{d-2}=\left( \frac{2i\hbar^2}{ma_1} \right)^{d-2} \left(\frac{\hbar^2 \Gamma^2\left(\frac{d}{2}\right)}{i\pi m(d-2)}\right).
\end{equation} 
The two columns of Table \ref{comparison1} match (and all odd dimensions match) when using the above conversion Eq. (\ref{eq:abar_a1_conversion}). We note that, when $d  =1, 3$ and $5$, the relationship between $\bar{a}_d$ and $a_1$ become
\begin{align}
\bar{a}_1=\frac{2}{a_1}, && \bar{a}_3= \frac{\hbar^4}{2m^2a_1}, &&\textrm{and}&& \bar{a}_5 =- \frac{\hbar^2}{ma_1}\sqrt[3]{\frac{3\hbar^2}{2m}}.
\label{aarelationship}
\end{align} 

As $d\to1$ the generalized pseudopotential (\ref{eq:generalPseudo_k0}) {\it prima facie} approaches zero. However, taking the $d\to1$ limit in the non-integer (``fractional'') derivative part of Eq. (\ref{eq:generalPseudo_k0}) and using the non-integer derivative (\ref{RLpower}), the $1d$ pseudopotential approaches the non-zero result shown in Table \ref{comparison1}. See Appendix \ref{subsec:C_1d} for the $1d$ derivation of $_{C_0}V_{k\to0}^{(1)} =-\frac{2}{\bar{a}_1} \, \delta^{(1)}(r)$. For even dimensions, the pseudopotential takes on a different form involving a logarithmic term inside the regularization operator. W\'odkiewicz did not present an explicit expression for the pseudopotential in the case of general even dimension. In the current work, we derived an expression for the W\'odkiewicz pseudopotential for all odd dimension and validated the equivalence of our non-integer operator. It may be useful for future study to produce a general even dimension expression for the W\'odkiewicz pseudopotential.


%
%


\section{Energy solution of two cold atoms in a harmonic trap in arbitrary dimension}\label{sec:5_two_cold}
In this section, we extend the Busch derivation in \cite{busch} of the energy for two cold atoms in a trap for $d=1, 2, 3$ to arbitrary dimension including non-integer. The Schr\"{o}dinger equation for the relative motion of two cold atoms in a harmonic trap is defined by:
\begin{equation}\label{eq:HO_Vd}
\left(H^{(d)}_{osc}+\hat{V}^{(d)}(r)\right)\Psi(r)=E^{(d)}\Psi(r).
\end{equation}
For the 3-dimensional case, Busch used $\sqrt{2}\pi a_0 \delta^{(3)}({\bf r})\frac{\partial}{\partial r}r$ in place of $\hat{V}^{(d)}(r)$ where $a_0$ is the $s$-wave scattering length (Eq. (3) of Ref.~\cite{busch}). Note that the Busch $a_0$ is a radius in whatever dimensional space being considered and is not related to the W\'odkiewicz coupling constant $a_d$. In this section, we solve the two trapped atom energy for arbitrary $d$ with our generalized pseudopotential $_{C_0}V_{k\to0}^{(d)}$ (Eq. \ref{eq:generalPseudo_k0}) in place of  $\hat{V}^{(d)}(r)$ in Eq. (\ref{eq:HO_Vd}). 
\subsection{Exact energy solution}
The energy for two cold trapped atoms with the generalized pseudopotential may be solved by expanding $\Psi$ in Eq. (\ref{eq:HO_Vd}) in terms of the $d$-dimensional harmonic oscillator basis wave functions, $\phi_n^{(d)}$, satisfying $H^{(d)}_{osc} \; \phi_n^{(d)}={E_n^{(d)}}\phi_n^{(d)}$. In Appendix \ref{appendix:HO}, we derive the solution of the $d$-dimensional harmonic oscillator, which is needed for the two-cold atom trap derivation below and Appendix \ref{appendixD}. Recalling our ultra-cold ($k\to0$) general $d$ pseudopotential from Eq. (\ref{eq:generalPseudo_k0}):
\begin{equation}
_{C_0}V_{k\to0}^{(d)} = \frac{\Omega(d)\bar{a}^{d-2}}{\Gamma(d-2)}\delta^{(d)}(r){\; _{C_0}D}_r^{d-2}r^{d-2},
\end{equation}
and expanding $\Psi(r)=\sum_{n=0}^\infty c_n\phi_n^{(d)}(\vec{r})$, Eq. (\ref{eq:HO_Vd}) becomes:
\begin{equation}
\label{eq:startBusch}
\begin{split}
&\sum_{n=0}^\infty c_n {E_n^{(d)}} \phi_n^{(d)} +\frac{\Omega(d)\bar{a}^{d-2}}{\Gamma(d-2)}\delta^{(d)}(r)\\
&\quad \times\Bigl[{_{C_0}D}_{r}^{d-2} r^{d-2}\sum_{n=0}^\infty c_n\phi_n^{(d)}
\Bigr]_{r\rightarrow 0}\\
&=\sum_{n=0}^\infty c_nE^{(d)}\phi_n^{(d)},
\end{split}
\end{equation}
where ${_{C_0}D}_{r}^{d-2}$ is the Caputo fractional derivative. We follow the integral procedure in Busch with special considerations for non-integer dimension (Appendix \ref{appendixD}) to reduce the energy to the following implicit equation for the energy
\begin{equation}
\label{eq:energyEquation_temp}
 	\frac{\pi(d-2) }{\Gamma^2( \frac{d}{2})}
\frac{\Gamma(-\frac{E^{(d)}}{2}+\frac{d}{4})}{\Gamma(-\frac{E^{(d)}}{2} + \frac{4-d}{4})} = \frac{-\sin\left(\frac{d\pi}{2}\right)}{\bar{a}_d^{d-2}} .
\end{equation}

In order to compare with the three dimensional results in Busch {\it et al.} we convert our $\bar{a}_d$ scattering length in $d$ dimensions to relative motion units (${a_o}$ used in Ref.~\cite{busch}):
\begin{equation} 
\label{eq:La0}
	\bar{a}^{d-2}_d = \frac{{a_o}^{d-2}}{2^{d/2}},
\end{equation}
where the $2^{d/2}$ arises from relative motion units in the $\delta^{(d)}({\bf r})$. Using this conversion our energy functional becomes
\begin{equation}
\label{eq:energyEquation}
 	\frac{\pi(d-2)}{\Gamma^2( \frac{d}{2})}
\frac{\Gamma(-\frac{E^{(d)}}{2}+\frac{d}{4})}{\Gamma(-\frac{E^{(d)}}{2} + \frac{4-d}{4})} = 
					\frac{-\sin\left(\frac{d\pi}{2}\right)} 
								{ {{a_o}}^{d-2}/2^{d/2} } .
\end{equation} 
\subsection{Approximate analytical solution for the energy}
Using a Taylor series expansion about $\bar{a}$ of Eq. (\ref{eq:energyEquation_temp}) (see Appendix \ref{appendixE}) and using the Eq. (\ref{eq:La0}) unit conversion ${a_o}$, we obtain a weakly interacting analytical solution for $E^{(d)}$: 
\begin{equation}\label{eq:approxEnergyEquation}
E^{(d)} \approx 2n +\frac{d}{2}
+ 
\frac{2\pi (d-2)}{\sin\left(\frac{d\pi}{2}\right)\Gamma^3\left(\frac{d}{2}\right)}
\frac
	{
	\Gamma \left(n+\frac{d}{2}\right)
	}
	{
	\Gamma \left(n+1\right)
	\Gamma \left(1-\frac{d}{2}\right)
	} { \frac{{a_o}^{d-2}}{2^{d/2}} }.
\end{equation}
When $d \to 3$,
\begin{equation}
E^{(3)} \approx 2n + \frac{3}{2} +
\frac
	{{2}^{3/2}}
	{\pi}
\frac
	{
	\Gamma \left(n+\frac{3}{2}\right)
	}
	{
	\Gamma \left(n+1\right)
	} {a_o},
\end{equation}
one can verify that the expression agrees with the Busch solution for $d=3$ (replacing the factorials in the binomial coefficient with gamma functions in Eq. (18) of Ref.~\cite{busch}).  
When $d \to 1$,
\begin{equation}
E^{(1)} \approx 2n + \frac{1}{2} -
\frac
	{2^{1/2}}
	{\pi}
\frac
	{
	\Gamma \left(n+\frac{1}{2}\right)
	}
	{
	\Gamma \left(n+1\right)
	} \frac{1}{{a_o}},
\end{equation}
one can verify that the expression agrees with the Busch solution for $d=1$ (replacing the factorials in the binomial coefficient with gamma functions in Eq. (20) of Ref.~\cite{busch}). 
 
To find energies near $d=2$, we expand Eq. (\ref{eq:energyEquation}) about $d=2$. To first order in $d-2$, Eq. (\ref{eq:energyEquation}) yields (see Eq.~\ref{eq:busch2d_derived})
\begin{equation}
\psi \left(\frac{1}{2} - \frac{E^{(2)}}{2} \right) = \ln \left( \frac{1}{{{a_o}}^2} \right),
\end{equation}
where $\psi(\cdot)$ is the logarithmic derivative of Euler's $\Gamma$-function. 

 \section{Discussion}\label{sec:discuss}
In this paper, we offer a generalization of the Fermi pseudopotential to all
values of dimension including non-integer values.  We thus extend previous
work on the Fermi pseudopotential generalizing this important concept
into new regimes.

Our approach circumvents the singularities at integer dimension by deriving
the generalization at non-integer values of the dimension. In the limit of
integer dimensions, we demonstrate that our generalized pseudopotential
agrees with previous integer d results. In addition, our generalization
includes energy dependence and simplifies earlier expressions.

Our approach necessitates the use of non-integer or fractional derivatives.
We review the fractional calculus required to derive our non-integer
pseudopotential in Appendix A and justify our choice of the particular 
non-integer
derivative used in the derivation.

In a further test, we applied our generalized pseudopotential to the problem 
of two cold atoms 
in a harmonic trap and obtained a general-$d$ relationship for the energy 
(Eqs. \ref{eq:energyEquation} and \ref{eq:approxEnergyEquation}) which agreed
with a previous solution in $d=1,2$, and $3$ dimensions\cite{busch}. 

For the two-atom trapped energy in arbitrary dimension, we used the ultra-cold 
($k \to 0$, i.e. energy-independent) limit of the generalized pseudopotential. 
In the strongly interacting regime, it may become important to include energy 
dependence in the interaction, for example, by using an energy-dependent 
scattering length \cite{blume,bolda}. Thus, an important future extension of 
our arbitrary-dimension trapped two-atom energy will be to obtain a solution 
that 
has energy dependence ($k>0$)\cite{Block_Holthaus}. 

An interesting application of current interest in atomic physics is the
creation of highly anisotropic trapping configurations of cold atoms yielding
quasi-one and quasi-two dimensional systems\cite{Ketterle01, bloch}. 
Quasi-one and-two dimensional systems can be created by using tight
confinement along one or two directions with weak confinement elsewhere.
These quasi one- and two- dimensional configurations
can be explored by tuning both the confinement and the interaction
parameters  to lead to 
unexplored physical regimes. Our work offers the interesting
approach of using a Fermi pseudopotential for  non-integer dimensions 
between one and two, characterized by the ratio of the different confinement
strengths.
	
While our pseudopotential 
is non-integer dimensional, it is valid arbitrarily close to integer dimension 
and gives correct limits for integer dimension. The limit for odd dimension is 
smooth. In even dimension, a term involving $\ln(r)$ arises in the 
series solution. This logarithmic behavior was also found in our $d=2$
two-cold-atom energy functional. Deriving an explicit 
regularization operator that removes the logarithmic singularities 
for even dimension remains a future project.

\appendix

\appendix\newpage\markboth{Appendix}{Appendix}
\renewcommand{\thesection}{\Alph{section}}
\numberwithin{equation}{section}

\section{Fractional calculus}\label{sec:frac_calc}

 We need to use a so-called ``fractional derivative'' operator. Fractional is
 the standard term used in the literature even though the order is valid 
beyond the domain of rationals~\cite{Lavoie76}. The fractional calculus 
approach has some advantages, but fractional derivatives must be applied 
carefully. There are various practical definitions of fractional order 
derivatives that depend on a lower integration limit (Eq. \ref{eq:fracDef} ). 
Each definition must behave the same in the limit of integer order; however, 
the properties of each derivative may lead to different, and sometimes
 counterintuitive, properties away from integer order. For our derivation of 
the non-integer dimension pseudopotential, we are primarily concerned with 
derivatives of powers of $r$ because of the series solution and the need to 
ensure that the general pseudopotential operator will converge near $r=0$. 
The Caputo definition with the Reimann-Liouville integration 
limit~\cite{Caputo67} helps ensure that our pseudopotential is well behaved 
for non-integer $d$ when $r \to 0$. However, our $d$ dimensional derivation 
of two-cold atoms in a trap also requires the fractional derivative of 
Gaussian functions ($e^{-r^2/2}$). We use the generalized Leibniz product 
rule to ensure consistent derivatives of Gaussians for all $d$, 
including non-integer.

We take advantage of fractional calculus to solve the $d$-dimensional pseudopotential and the energy for two-cold atoms in a trap in $d$ dimensions. Fractional calculus has also been used in domains to describe systems such as anomalous diffusion~\cite{diffusion:82} or the dynamics of neuron cluster response that may adapt on multiple time scales~\cite{frac_neuro}.

Most definitions of fractional derivatives differ by the choice of the lower limit of integration in the definition. For a general lower integration limit $x_0$, the definition of a fractional derivative, ${x_0}D^{\alpha}_{x}$, of order $\alpha$ with respect to $x$ is given as follows (see~\cite{Lavoie76}):
\begin{equation}\label{eq:fracDef} 
 _{x_{\mathstrut 0}}D^{\alpha}_{x}f(x)=\frac{1}{\Gamma(-\alpha)}\int^{x}_{x_{\mathstrut 0}}(x-x')^{-\alpha-1}f(x')dx'.
\end{equation}
``Fractional" is the standard term used in the literature even though $\alpha$ is valid beyond the domain of rationals.  The order $\alpha$ may be real or complex and the arbitrary lower-limit of integration $x_0$ arises from iterated integration.  Any selections for $x_0$ will yield the results of conventional calculus when $\alpha$ is integer, but, in general, different selections for the lower limit $x_0$ will result in the emergence of different properties away from integer $\alpha$.  There are many fractional derivatives defined in the literature, the most common being Reimann-Liouville ($x_0=0$), Weyl ($x_0 \to -\infty$), and Caputo~\cite{podlubny}. In general, fractional differentiation is non-local and collapses to a local calculation for integer-order derivatives.  For fractional differentiation in the time domain, this non-local behavior can be interpreted as taking into account the history of the function. The value of the lower bound determines the duration of the memory of the differentiation.  

To show the effect of the lower integration limit $x_0$, one may show from Eq. (\ref{eq:fracDef}) the following formula for the fractional derivative of a power, $x^\beta$:
\begin{equation}\label{eq:generalPower}
\begin{split}
& _{x_{\mathstrut 0}}D^{\alpha}_{x}x^{\beta} = \frac{\Gamma(\beta+1)}{\Gamma(\beta-\alpha+1)}x^{\beta-\alpha}\\
&\quad -\frac{x_0^{\beta+1}}{(\beta+1)\Gamma(-\alpha)}x^{-\alpha-1} {_2F_{1}}(1+\alpha,1+\beta;2+\beta;\frac{x_0}{x}),
\end{split}
\end{equation}
where $_2F_1(\cdot,\cdot;\cdot;\cdot)$ is the Gaussian hypergeometric function. The first term on the right-hand side of Eq. (\ref{eq:generalPower}) would be the only term if factorials were simply replaced by gamma functions in integer-order differentiation. The second term goes to 0 when the order $\alpha$ is integer, but when $\alpha$ is not integer, the second term depends on the lower integration limit $x_0$ from Eq. (\ref{eq:fracDef}). Letting $x_0=0$ in the definition is a reasonable choice (Reimann-Liouville), but other familiar properties may change away from integer order differentiation, such as the derivative of $e^x$ not being itself, the derivative of a constant not being zero, or the usual product rule not being valid. In the remainder of this section, we describe some of the issues of common fractional derivative definitions and the rationale behind our definition.       


\subsection{Riemann-Liouville derivative}
Selecting $x_0 = 0$ yields the Riemann-Liouville (RL) derivative, defined and denoted as follows:
\begin{equation}
_0D^\alpha_xf(x)=\frac{1}{\Gamma(-\alpha)}\int^x_0\left(x-x'\right)^{-\alpha-1}f\left(x'\right)dx'.
\end{equation}
For RL dervatives, setting $x_0=0$ in Eq. (\ref{eq:generalPower}), we see that the derivative of a power of $x$ has the following form:
\begin{equation}
\label{RLpower}
_0D^\alpha_x x^\beta = \frac{\Gamma(\beta+1)}{\Gamma(\beta-\alpha+1)}x^{\beta-\alpha},
\end{equation}
which is a simple extension of the pattern that emerges from differentiation of integer orders. From the above equation, it can be seen that the RL derivative of a constant is, in general, nonzero, which poses fundamental problems for our series solution and the $r\to0$ limit of the pseudopotential derivation. However, the Caputo derivative addresses this problem, next.  

\subsection{Caputo derivative}
An alternative formulation of the fractional derivative that retains the utility of the RL derivative with respect to power functions while allowing the additional property that the derivative of a constant is zero is the Caputo fractional derivative (CFD) with the RL lower integration limit $x_0 = 0$.  In general, the lower limit in the CFD may be $x_0$ as above, but we focus on the $x_0=0$ RL limit because of our extensive use of powers in the series solution of the Schr\"odinger equation. Proposed by Caputo for the theory of viscoelasticity~\cite{Caputo67}, the CFD is defined as follows:
\begin{equation}\label{eq:caputoDef}
_{C_0}D^\alpha_x f(x) = \frac{1}{\Gamma(m-\alpha)} \int^x_0 \left(x-x'\right)^{m-\alpha-1} f^{(m)}\left(x'\right)dx'
\end{equation}
with $m = \lceil\alpha\rceil$ (the ceiling of $\alpha$). The subscript $C_0$ is meant to indicate the Caputo derivative and the lower integration limit $x_0=0$. The CFD is defined for $\alpha\in\mathbb{R}^+$ (the positive real numbers), though it can be extended to complex orders. This CFD possesses the RL pattern for powers (Eq. \ref{RLpower}) with the additional property that
\begin{equation}
\label{CaputoPower}
_{C_0}D^\alpha_x x^\beta=0 \textrm{ for } \beta\in\mathbb{Z}^+ \textrm{ and }\alpha>\beta,
\end{equation}
where $\mathbb{Z}^+$ is the set of all positive integers. The property in Eq. (\ref{CaputoPower}) is essential for the series solution approach we adopt in Sec.~\ref{sec:3.2} and Appendix \ref{A_SchrodingerSolDeriv}. This property is satisfied by Eq. (\ref{eq:caputoDef}) because the integer part of a real-order derivative is applied first ($f^{(m)}$) and then the remaining non-integer ($m-\alpha$) derivative is taken. 

For the derivation of the implicit function for the energy of two cold atoms in a trap in fractional dimension (Section \ref{sec:5_two_cold} and Appendix \ref{appendixD}), we take the fractional derivative of Gaussian functions from the harmonic oscillator basis functions ($e^{-r^2/2}$). To calculate these derivatives, we use the Leibniz product rule for Caputo derivatives, which is extended to fractional order~\cite{ishteva_thesis}:
\begin{equation}\label{eq:general_Leibniz}
\begin{split}
&D_{r}^\eta(f(r),g(r))\\
& = \sum_{k=0}^\infty\binom{\eta}{k}(D_{r}^{\eta-k}f(r))g^{(k)}(r) - \sum_{k=0}^{m-1}\frac{r^{k-\eta}}{\Gamma(k+1-\eta)}(fg)^{(k)}(0),
\end{split}
\end{equation}
where $m = \lceil\eta\rceil$. We also use the generalized Leibniz rule to take the $d\to1$ limit of our generalized psuedopotential in Appendix \ref{subsec:C_1d}.

\section{Derivation of pseudopotential for $k \to 0$}\label{small_k}
  Assuming that $B$ is finite and independent of $r$, we can construct the 
following from Eq.~(\ref{boundaryCondition}) when $r$ is small:
\begin{equation}
\begin{split}
\bigl[{\; _{C_0}D}_r^{d-2}r^{d-2} \psi(r)\bigr]_{r\to0}\\ 
&=B{\; _{C_0}D}_r^{d-2} \bigl(r^{d-2}-\bar{a}^{d-2}\bigr)\\  
&= B{\; _{C_0}D}_r^{d-2} r^{d-2} = B \Gamma(d-1),
\end{split}
\end{equation}
where we allow non-integer $d$ and adopt Caputo's definition 
(Eq.~\ref{eq:caputoDef}) for non-integer (``fractional'') order 
derivatives, ${_{C_0}D}_r^{d-2}$, which follows the Reimann-Liouville 
formulas for differentiating power functions (Eq. \ref{RLpower}) with the additional important property of Eq. (\ref{CaputoPower}).
It follows that
\begin{equation}
\label{mychi}
B = \frac{1}{\Gamma(d-1)}\left[{\; _{C_0}D}_r^{d-2}r^{d-2} \psi(r)\right]_{r\to0} .
\end{equation}
On the other hand, we can also construct the following from Eq.~(\ref{boundaryCondition}), as $r \to 0$:
\begin{equation}
\label{boundaryLimits}
r^{d-1}\frac{d\psi}{dr} \to r^{d-1}B (d-2)\frac{\bar{a}^{d-2}}{r^{d-1}} = B (d-2)\bar{a}^{d-2}.
\end{equation}
We now integrate both sides of Eq. (\ref{boundaryLimits}) over the 
full solid angle. The left side of Eq. (\ref{boundaryLimits}) becomes 
(see Ref.~\cite{Blatt} p. 75)
\begin{equation}\label{eq:vol_integral}
\int r^{d-1}\frac{d\psi}{dr} d\Omega = \int (\nabla_d \psi)\cdot \textbf{n} \,dS = \int \left(\nabla^2_d \psi\right)dV,
\end{equation}
where \textbf{n} is a unit vector normal to the surface of the $d$-dimension
 hypersphere, and the right side of Eq. (\ref{boundaryLimits}) 
becomes \begin{equation}\label{eq:vol_int_rhs}
\int B (d-2)\bar{a}^{d-2} d\Omega =B (d-2)\bar{a}^{d-2}\int  d\Omega = B (d-2)\bar{a}^{d-2} \Omega(d),
\end{equation}
where $\Omega(d)=2\pi^{d/2}/\Gamma(d/2)$. Equating the right hand sides 
of Eqs. (\ref{eq:vol_integral}) and (\ref{eq:vol_int_rhs}), we obtain 
\begin{equation}
\int \left(\nabla^2_d \psi\right)dV = B (d-2)\bar{a}^{d-2} \Omega(d),
\end{equation}
and interposing an integral of a delta function on the right side, we 
can equate integrands: 
\begin{equation}\label{eq:interpose}
\int \left(\nabla^2_d \psi\right)dV = \int B (d-2)\bar{a}^{d-2} \Omega(d) \delta^{(d)}(r) dV,
\end{equation}
Thus, as $r \to 0$, substituting $B$ from Eq. (\ref{mychi}) into 
Eq. (\ref{eq:interpose}), we obtain the limit of the integrand:
\begin{equation}
\label{prePseudoSmallk}
\nabla^2_d \psi \to \frac{\Omega(d)(d-2) \bar{a}^{d-2}}{\Gamma(d-1)}\delta^{(d)}(r){\; _{C_0}D}_r^{d-2}r^{d-2} \psi.
\end{equation}
Simplifying the operator on the right side of Eq.~(\ref{prePseudoSmallkA}), 
we obtain the pseudopotential for small $k$,
\begin{equation}
\label{pseudoSmallkA}
_{C_0}V_{k\to0}^{(d)} (r) = \frac{\Omega(d) \bar{a}^{d-2}}{\Gamma(d-2)}\delta^{(d)}(r){\; _{C_0}D}_r^{d-2}r^{d-2}.
\end{equation}
Eq. (\ref{pseudoSmallk}) will be the limiting case ($k\to0$) of the more 
general $k$-dependent pseudopotential (Eq. \ref{pseudoPotential}) in the 
next section, and it will be the starting point for the ultra-cold 
two-trapped-atom derivation in Section \ref{sec:5_two_cold}.

\section{Fractional-Frobenius method for series solution of the wave function}
\label{A_SchrodingerSolDeriv}
Before constructing the pseudopotential we first perform a series solution to identify the singularities to be removed. We use the Frobenius method to construct the series solution, but in our derivation of the pseudopotential we assume that $d$ is fractional. Non-integer $d$ avoids the $\ln(r)$ singularities that arise for all integer $d$ except for $1$ and $3$, and because $d$ is allowed to be arbitrarily close to integer, the pseudopotential can be calculated for integer $d$ as a limit.  

We rewrite the Laplacian of the relative motion for $r>\bar{a}$ (Eq. (\ref{Schrodinger})):
\begin{equation}
\label{eq:A1}
\left(\frac{1}{r^{d-1}}\frac{d}{dr}r^{d-1}\frac{d}{dr}+k^2\right)\psi(r)=0,
\end{equation}
as a second ordered linear differential equation that can be solved by the Frobenius method:
\begin{equation}\label{eq:A2}
\psi''+\frac{d-1}{r}\psi'+k^2\psi=0.
\end{equation}
Substituting $\psi(r)=\sum_{\lambda=0}^\infty\alpha_\lambda r^{j+\lambda}$ and simplifying, we are left with 
\begin{equation}
\sum_{\lambda=0}^\infty\alpha_\lambda(j+\lambda)(j+\lambda+d-2)r^{j+\lambda-2}+k^2\sum_{\lambda=0}^\infty\alpha_\lambda r^{j+\lambda},
\end{equation}
yielding the indicial equation:
\begin{equation}
j(j+d-2)=0
\end{equation}
and a recurrence relation:
\begin{equation}
\alpha_{n+2}=\frac{-k^2\alpha_n}{(n+2+j)(n+d+j)}.
\end{equation}
The roots of the indicial equation are $j=0$ and $j =2-d$, where the first root yields the regular solution:
\begin{equation}
\psi_{reg}(r)=\sum_{n=0}^\infty\alpha_nr^n,
\end{equation}
with a recursive coefficient function:
\begin{equation}
\alpha_{n+2}=\alpha_n\cdot\frac{-k^2}{(n+2)(n+d)}
\end{equation}
where $\alpha_1=0$, which sets all $\alpha_{odd} = 0$, and $\alpha_0$ is an arbitrary, non-zero constant. By simple algebraic manipulation, we can rewrite the formula for $\alpha_n$:
\begin{equation}
\label{eq:alphan}
\alpha_{2n} = \alpha_0\cdot\frac{(-1)^{n}k^{2n}(d-2)!!}{(2n)!!(2n+d-2)!!}, \quad n\in\mathbb{Z}^+.
\end{equation}
With the second indicial root $j=2-d$, the second solution becomes
\begin{equation}
\label{eq:A9}
\psi_{irreg}(r)=c\ln(r)\psi_{reg}+\frac{1}{r^{d-2}}\sum_{n=0}^\infty\beta_nr^n.
\end{equation}

For completeness, we solve for the coefficients $c$ and $\beta_n$ for all cases of $d$. However, since our approach uses ``fractional'' $d$ we only need case $2.$) $d \notin \mathbb{Z}$. In this case, $c=0$ and the $\ln(r)$ term is removed (see discussion near Eqs.~\ref{eq:A11} and \ref{eq:betan_appendixA}). The non-integer-$d$ series solution can be made arbitrarily close to integer, and we will find that the limit for odd $d$ is smooth while the even $d$ limit must be handled more carefully.

We now solve for the coefficients $c$ and $\beta_n$ by substituting the irregular solution $\psi_{irreg}(r)$ from Eq. (\ref{eq:A9}) into the original equation Eq. (\ref{eq:A2}): 
\begin{equation}
\label{mainEq1}
\begin{split}
&\sum_{n=0}^\infty c \alpha_n(2n-2+d)r^{n-2}+\beta_1(3-d)r^{1-d}\\
&\quad +\sum_{n=0}^\infty
\bigl[\beta_{n+2}(n+4-d)(n+2)\\
& \quad + k^2\beta_n\bigr] r^{n+2-d}=0
\end{split}
\end{equation}
Taking note that the first series is a sum of powers of $r$ starting at $r^{-2}$ and that the other terms are series of powers of $r$ that depend on dimension $d$, we solve for the constants $c$ and $\beta_n$ for the following cases. 

\begin{enumerate}[1.)]

\item $d \in \mathbb{Z}$:
	\begin{enumerate}
\item $d = 2$:
		
\noindent Eq. \ref{mainEq1} can be rewritten as:
\begin{equation}
\begin{split}
&\sum_{n=1}^\infty c \alpha_n(2n)r^{n-2}+\beta_1r^{-1}\\
&+\sum_{n=0}^\infty\left[\beta_{n+2}(n+2)^2 + k^2\beta_n\right] r^{n}=0
\end{split}
\end{equation}
\noindent or
\begin{equation}
\sum_{n=2}^\infty \bigl[2nc \alpha_{n} + \beta_{n}n^2 + k^2\beta_{n-2}\bigr]r^{n-2}+(2c\alpha_1+\beta_1)r^{-1}=0,
\end{equation}
\noindent which implies $\beta_1 = 0$ (since $\alpha_1 = 0$) and $2nc \alpha_{n} + \beta_{n}n^2 + k^2\beta_{n-2}=0$. Hence, 
\begin{equation}
 c = -\frac{k^2\beta_{n-2}+\beta_{n}n^2}{2n\alpha_{n}} = -\frac{3\beta_3}{2\alpha_3}.
\end{equation}

\item $d = 3$:
		
\noindent Eq. (\ref{mainEq1}) can now be rewritten as:
\begin{equation}
\label{d=3}
\begin{split}
&\sum_{n=0}^\infty c \alpha_n(2n+1)r^{n-2}+\sum_{n=0}^\infty\bigl[\beta_{n+2}(n+1)(n+2)\\
&\quad + k^2\beta_n\bigr] r^{n-1}=0
\end{split}
\end{equation}
\noindent The first sum in Eq. (\ref{mainEq1}) requires $c=0$; the second term is automatically 0, making $\beta_0$ and $\beta_1$ arbitrary; and the third term yields the recursive coefficient function: 
\begin{equation}
\beta_{n+2}=\frac{-k^2\beta_n}{(n+1)(n+2)},
\end{equation}
\noindent or
\begin{equation}
\beta_{n}=\beta_{\bar{n}_2}\frac{(-1)^{\lfloor\frac{n-1}{2}\rfloor}k^n}{n!!(n-1)!!},
\end{equation}
\noindent where $\bar{n}_2$ denotes the common residue of $n$ mod 2.

\item $d \geq 4$: 

Eq. \ref{mainEq1} can be rewritten as:
\begin{multline}
\label{eq:dgeq4}
\beta_1(3-d)r^{-1} + \sum_{n=2}^{d-3}(n \beta_n + \beta_{n-2}k^2)r^{n-d} + \\
\sum_{n =d-2}^\infty\bigl[c \alpha_{n-d+2}(2n+2-d) + \beta_{n}n(n+2-d) \\
+ k^2\beta_{n-2}\bigr] r^{n-d}=0
\end{multline}
		
\noindent The first term implies $\beta_1=0$. The second term reveals the recurrence relation:
\begin{equation}
\label{eq:betaRelation:dgeq4}
\beta_{n}=\frac{-k^2\beta_{n-2}}{n}, \quad n = 2,3,\dots, d-3
\end{equation}
\noindent and hence $\beta_{2j+1}=0$ for $j < d/2-2$. The third term requires
\begin{equation}
c=-\frac{\beta_n n(n+2-d)+k^2\beta_{n-2}}{\alpha_{n+2-d}(2n+2-d)},\quad n = d-3, d-2,\dots.
\end{equation}
\noindent We note that in the special case of $d = 4$, the second term in Eq. (\ref{eq:dgeq4}) disappears, and we do not have the second recurrence relation of $\beta$ described in Eq. (\ref{eq:betaRelation:dgeq4}). Therefore, in this case, most values of the $\beta$'s (except $\beta_1$) are arbitrary. 

\item Otherwise:

\noindent When $d\leq1$, we can rewrite Eq. \ref{mainEq1} as:
\begin{multline}
\sum_{n=0}^{3-d} c \alpha_n(2n-2+d)r^{n-2}+\beta_1(3-d)r^{1-d}+\\
\quad\sum_{n=4-d}^\infty\bigl[c \alpha_n(2n-2+d)\\
 + \beta_{n+d-2}n(n+d-2) + k^2\beta_{n+d-4}\bigr] r^{n-2}=0,
\end{multline}
\noindent which leads to $c=0$, $\beta_1=0$, and $\beta_0$ is arbitrary, yielding the recursive coefficient function from Eq. (A5): 
\begin{equation}
\beta_{n}=\frac{-k^2\beta_{n-2}}{n(n+2-d)}, \quad n = 2,3,\dots
\end{equation}
\noindent which sets all $\beta_{odd}$ to zero, and generalizes $\beta_n$ to
\begin{equation}
\beta_n=\beta_0\cdot\frac{(-1)^{n/2}k^n(d-2)!!}{n!!(n+2-d)!!}, \quad n \in \mathbb{Z}_{even}.
\end{equation}
\hspace{2mm}
\end{enumerate}

\item $d \notin \mathbb{Z}$:

The first sum in Eq.~(\ref{mainEq1}) does not have common powers of $r$ with the second and third term, forcing $c = 0$. Also, when $d \notin \mathbb{Z}$, $3-d\neq0$, which implies $\beta_1 = 0$. The last sum yields the recursive relation:
\begin{equation}
\label{eq:A11}
\beta_{n+2}=\frac{-k^2\beta_n}{(n+4-d)(n+2)},
\end{equation}
or
\begin{equation}
\label{eq:betan_appendixA}
\beta_{2n}=\beta_{0}\cdot\frac{(-1)^{n}k^{2n}(2-d)!!}{(2n)!!(2n+2-d)!!}, \quad  n\in\mathbb{Z}^+.
\end{equation}
\end{enumerate}

In conclusion, for fractional dimension $d\notin \mathbb{Z}$, the solution to the Schr\"{o}dinger equation has the form:
\begin{equation}
\psi(r)=\sum_{n=0}^\infty\alpha_nr^n+\frac{1}{r^{d-2}}\sum_{n=0}^\infty\beta_n r^n.
\end{equation}
Applying the boundary condition of the Schr\"odinger equation, we have:
\begin{align*}
0 & = \sum_{n=0}^\infty \alpha_{2n} \bar{a}^{2n} + \sum_{n=0}^\infty \beta_{2n} \bar{a}^{2n+2-d}\\
& = \alpha_0\sum_{n=0}^\infty \frac{(-1)^{n}k^{2n}(d-2)!!\bar{a}^{2n}}{(2n)!!(2n+d-2)!!}\\
& + \beta_0 \sum_{n=0}^\infty \frac{(-1)^{n}k^{2n}(2-d)!!\bar{a}^{2n+2-d}}{(2n)!!(2n+2-d)!!}.
\end{align*}
Hence,
\begin{equation}
\label{alpha0beta0}
\begin{split}
\alpha_0&= - \beta_0 
\frac
	{\sum_{n=0}^\infty \frac{(-1)^{n}k^{2n}(2-d)!!\bar{a}^{2n+2-d}}{(2n)!!(2n+2-d)!!}}
	{\sum_{n=0}^\infty \frac{(-1)^{n}k^{2n}(d-2)!!\bar{a}^{2n}}{(2n)!!(2n+d-2)!!}}\\
&=  - 
\frac{\beta_0 }{\bar{a}^{d-2}}
\frac
	{_0F_1\Bigl(;2-\frac{d}{2}; - \frac{k^2\bar{a}^2}{4}\Bigr)}
	{ {_0F}_1\left(;\frac{d}{2}; - \frac{k^2\bar{a}^2}{4}\right)}.
\end{split}
\end{equation}
where $_0F_1(;\cdot; \cdot)$ is the confluent hypergeometric function \cite{handbookMathFuncs}.

\section{Integration over $\epsilon$ sphere for general $d$ pseudopotential}\label{appendixB}
Here we provide details of the derivation of Eq. (\ref{eq:int_SE_beta0}) in Section \ref{sec:3.2}. This involves the integration of the Schr\"odinger equation (Eq. \ref{Schrodinger}) about a $d$-dimensional sphere with infinitesimal radius $\epsilon$. We begin with the non-integer-$d$ solution ($c=0$, Eq.\ref{generalSolForm}):
\begin{equation}\label{generalSolFormB}
\psi(r)=\sum_{n=0}^\infty\alpha_{2n}r^{2n}+\frac{1}{r^{d-2}}\sum_{n=0}^\infty\beta_{2n}r^{2n},
\end{equation}
The regular part of $\psi$ (right hand side of Eq.\ref{generalSolFormB}), $\psi_\textrm{reg} =\sum_{n=0}^\infty\alpha_{2n}r^{2n}$, does not contain singularities, so as $r \to 0$, $\left(\nabla^2_d + k^2\right)\psi_\textrm{reg} \to 0$. Thus, we only need to consider the irregular solution (left hand side of Eq.\ref{generalSolFormB}) when integrating the Schr\"odinger equation over the infinitesimal sphere:
\begin{align}
\int\left(\nabla^2_d + k^2\right)\psi dV & =
\int\left(\nabla^2_d + k^2\right)\psi_\textrm{irreg} dV\\& = 
\int\nabla^2_d\psi_\textrm{irreg} dV + \int  k^2 \psi_\textrm{irreg}dV \label{eps_integral1}
\end{align}
The integral involving $k^2$ is zero:
\begin{align}
\left[\int  k^2 \psi_\textrm{irreg}dV\right]_{\epsilon \to 0} &= 
\left[\int  k^2 \left(\sum_{n=0}^\infty\beta_nr^{2n-d+2}\right)dV\right]_{\epsilon \to 0}  \\&= 
\left[\int_0^{\epsilon}  k^2 \left(\sum_{n=0}^\infty\beta_nr^{2n-d+2}\right)r^{d-1}dr\right]_{\epsilon \to 0} \\&=
\left[\int_0^{\epsilon}  k^2 \left(\sum_{n=0}^\infty\beta_nr^{2n+1}\right)dr\right]_{\epsilon \to 0} \\&=
\left[k^2 \sum_{n=0}^\infty\frac{\beta_n}{2n+2}\epsilon^{2n+2} \right]_{\epsilon \to 0}  \\&=
0.
\end{align}
Returning to Eq. (\ref{eps_integral1}): 
\begin{align}
\int\left(\nabla^2_d + k^2\right)\psi dV & = \int\nabla^2_d\psi_\textrm{irreg} dV + \cancelto{0}{\int  k^2 \psi_\textrm{irreg}dV}\\
& = \int\nabla^2_d\left(\sum_{n=0}^\infty\beta_{2n}r^{2n-d+2}\right) dV\\
& = \int dS \cdot \nabla_d\left(\sum_{n=0}^\infty\beta_{2n}r^{2n-d+2}\right) \\
& = \Omega(d) \epsilon^{d-1}\sum_{n=0}^\infty\beta_{2n}(2n-d+2)\epsilon^{2n-d+1}\\ 
& = \Omega(d) \sum_{n=0}^\infty\beta_{2n}(2n-d+2)\epsilon^{2n};
\end{align}
We take the limit $\epsilon\to 0$ and interpose the integral of a delta function:  
\begin{align}
\int\left(\nabla^2_d + k^2\right)\psi dV &\to \Omega(d) \beta_{0}(2-d)\\
\int\left(\nabla^2_d + k^2\right)\psi dV &= \int \Omega(d) \beta_{0}(2-d) \delta^{(d)}(r) dV. 
\end{align}
Equating the integrands gives the following and, thence, Eq. (\ref{eq:int_SE_beta0})
\begin{equation}
\label{pseudoBeta0}
\left(\nabla^2_d + k^2\right)\psi = \Omega(d) \beta_{0}(2-d) \delta^{(d)}(r).
\end{equation}

\section{Relationship between the generalized pseudopotential and the W\'odkiewicz Fermi pseudopotential (Green's function approach) regularizing factor $\gamma_{2n+1}$ and coupling constant $a_d$}\label{appendixC}
\subsection{The regularization factor}\label{AppendixC1}
At first glance, Eq. (\ref{eq:generalPseudo_k0}) derived by the non-integer calculus approach appears not to agree with the regularizing operator derived in Ref.~\cite{wod} by W\'odkiewicz for odd dimension $(d=2n+1)$ using the Green's function approach (Eq. (5.4) in \cite{wod}), 
\begin{equation}\label{eq:wod1}
\hat{R}_{W}^{(2n+1)}=\gamma_{2n+1}\frac{\partial^{2n-1}}{\partial r^{2n-1}}r^{2n-1},
\end{equation}
because of  the complicated looking analytical form of the regularization coefficient (Eq. (5.5) in \cite{wod}):
\begin{equation}\label{eq:wod2}
\gamma_{2n+1}=\frac{\pi^{-1/2}\Gamma\left(\frac{1}{2}-n\right)}{\sum_{l=0}^{n-1}(-1)^{l+n}2^{-l+n}\frac{(n-1+l)!(2n-1)!}{l!(n-1-l)!(n+l)!}}.
\end{equation}
However, as we show next, the result of our non-integer calculus approach (Eq. \ref{eq:generalPseudo_k0}) and the Green's function calculation for odd integer dimension (Eqs. \ref{eq:wod1} and {eq:wod2}) are in agreement, and we obtain the simplifying result for Eq. (\ref{eq:wod1}) taken from Ref.~\cite{wod} for positive integer $n$: 
\begin{equation}\label{eq:wod_connection}
\gamma_{2n+1}=\frac{1}{\Gamma(2n)}=\frac{1}{\Gamma(d-1)}.
\end{equation}
We now verify by direct means the relationship in Eq. (\ref{eq:wod_connection}) for $n\in \mathbb{Z}$ , which we predicted based on consistency between the non-integer calculus and Green's function derivations of the regularizing operator. We first note that the formula in Eq. (\ref{eq:wod2}) is obfuscated by the summation in the denominator, but this sum can be simplified to $(-1)^n2^{2n-1}(n-1)!$ or $(-1)^n2^{2n-1}\Gamma(n)$, which we use to replace the summation in Eq. (\ref{eq:wod2}): 
\begin{equation}\label{eq:wod_temp1}
\gamma_{2n+1}=\frac{\Gamma\left(\frac{1}{2}-n\right)}{\sqrt{\pi}(-1)^n2^{2n-1}\Gamma(n)}.
\end{equation}
In the denominator of Eq. (\ref{eq:wod_temp1}), we use the identity $\Gamma(z)\Gamma\left(z+\frac{1}{2}\right)=2^{1-2z}\sqrt{\pi}\Gamma(2z)$, which when rearranged we may replace $2^{2n-1}\Gamma(n)$ in the denominator of Eq. (\ref{eq:wod_temp1}) to give
\begin{equation}\label{eq:wod_temp2}
\gamma_{2n+1}=\frac{\Gamma\left(\frac{1}{2}-n\right)\Gamma\left(\frac{1}{2}+n\right)}{\pi(-1)^n\Gamma(2n)}.
\end{equation}
Substitution of $z=n+\frac{1}{2}$ into the identity $\Gamma(1-z)\Gamma(z)=\frac{\pi}{\sin(\pi z)}$ gives $\Gamma\left(\frac{1}{2}-n\right)\Gamma\left(\frac{1}{2}+n\right)=\frac{\pi}{\sin\left(\pi\left(\frac{1}{2}+n\right)\right)}$, and substituting this into Eq. (\ref{eq:wod_temp2}) yields
\begin{equation}\label{eq:wod_temp3}
\gamma_{2n+1}=\frac{1}{\Gamma(2n)}\frac{1}{(-1)^n \sin\left(\pi\left(\frac{1}{2}+n\right)\right)}
\end{equation}
One can now see from Eq. (\ref{eq:wod_temp3}) that Eq. (\ref{eq:wod_connection}) holds for positive and negative integers $n$ and $0$, since $(-1)^n \sin\pi\left(n+\frac{1}{2}\right)=1$ for $n\in Z$, and thus we demonstrate the consistency between the Green's function and non-integer calculus approaches.  The positive and negative domains for $n$ correspond to positive and negative odd dimensionality, respectively. The Eq. (\ref{eq:wod_connection}) coefficient goes to $0$ for negative (integer) $n$ or, equivalently, negative integer dimensionality. We show in Section \ref{subsec:C_1d} below that the generalized pseudopotential is finite and non-zero when $d\to1$. 

\subsection{The general (odd $d>1$) expression for W\'odkiewicz pseudopotential}\label{AppendixC2}
W\'odkiewicz derived the Fermi pseudopotential for $d = 1, 3$ and 5. We use the same procedure to obtain the general result for all odd $d$. Starting with the formula for the scattering amplitude [Corrected version of Eq. (3.1), Ref.~\cite{wod}, which contains a typographical error as we note in Appendix \ref{typoAppendix}]:
\begin{equation}f_k^{(d)}=-\frac{i^{-n} k^{n-1} (2 n-1)\text{!!} }{k^{2 n-1}+i  2^{n-1} \frac{\pi ^n (2 n-1)\text{!!}\hbar}{  a_{2 n+1} \alpha}}
\end{equation} 
we obtain the solution for the single pole of $f_k^{(d)}$ in the physical $k$ half plane:
\begin{equation}
k^{2 n-1}=-i  2^{n-1} \frac{\pi ^n (2 n-1)\text{!!}\hbar}{  a_{2 n+1} \alpha}.
\end{equation} 
Now, using the following $d=1$ single bound state energy [Eq. (3.3) in Ref.~\cite{wod}]:
\begin{equation}
E_0 = -\frac{m}{2\hbar^2}a_1^2
\end{equation} 
and setting the energy constant across dimensions, we have:
\begin{align}
&& E & =  E_0\\
\Rightarrow && \frac{\hbar^2k^2}{2m} &= -\frac{m}{2\hbar^2}a_1^2\\
\Rightarrow && k^2 &= -\frac{m^2}{\hbar^4}a_1^2\\
\Rightarrow && k &= \pm\frac{im}{\hbar^2}a_1\\
\Rightarrow && k^{2n-1} &= \pm\left(\frac{im}{\hbar^2}a_1\right)^{2n-1}\\
\Rightarrow && \frac{ -i  2^{n-1} \pi ^n (2 n-1)\text{!!}\hbar}{  a_{2 n+1} \alpha} &= \pm\frac{i^{2n-1}m^{2n-1}}{\hbar^{2(2n-1)}}a_1^{2n-1}
\hspace{20mm}
\end{align} 
from which we can clearly solve for $a_{2n+1}$:
\begin{equation}
a_{2 n+1}= \frac{(-1)^n\hbar^{4n}(2\pi)^n(2n-1)!!}{a_1^{2n-1}m^{2n}}, 
\end{equation}
or, since $d = 2n+1$,
\begin{equation}\label{eq:appendC_W_ad}
a_{d}=\frac{(-2 \pi )^{\frac{d-1}{2}} (d-2)\text{!!} \hbar^{2d-2}}{a_1^{d-2} m^{d-1}}.
\end{equation} 
We can now substitute our general formula above for the strength or ``area of potential,'' $a_d$, into the following relation [Eq. (1.2) in \cite{wod}] 
\begin{equation}
V^{(d)}_{W}(r) = -a_d\delta^{(d)}(r)\hat R_d 
\end{equation} 
and we obtain the general formula of W\'odkiewicz Fermi pseudopotential for odd $d>1$:
\begin{equation}
\label{eq:W1}
\begin{split}
&V^{(d)}_{W}(r)\\
& = -\frac{\hbar^{2d-2}(-2 \pi )^{\frac{d-1}{2}} (d-2)\text{!!} }{m^{d-1}a_1^{d-2} } \delta ^{(d)}(r) \frac{1}{\Gamma (d-1) }\frac{\partial^{d-2}}{\partial r^{d-2}} r^{d-2}\\ 
&= -\frac{\hbar^{2d-2}(-2 \pi )^{\frac{d-1}{2}} (d-4)\text{!!} }{m^{d-1}a_1^{d-2} } \delta ^{(d)}(r) \frac{1}{\Gamma (d-2)} \frac{\partial^{d-2}}{\partial r^{d-2}} r^{d-2}. 
\end{split}
\end{equation} 
We rewrote $V^{(d)}_{W}(r)$ in the second form so that it more closely resembles our general-dimension pseudopotential.

\subsection{$d\to1$ comparison of the W\'odkiewicz and generalized pseudopotentials}\label{subsec:C_1d}
When $d = 1$ in the W\'odkiewicz pseudopotential, the regularization factor $\frac{1}{\Gamma (d-1) }\frac{\partial^{d-2}}{\partial r^{d-2}} r^{d-2}$ is not required due to the lack of singularities in the limit $r\to 0$ of the Green's functions~\cite{wod}; hence, dropping the regularizing operator in Eq. (\ref{eq:W1}) gives
\begin{equation}
V^{(1)}_{W}(r) =-\frac{\hbar^{2-2}(-2 \pi )^{\frac{1-1}{2}} (1-2)\text{!!} }{m^{1-1}a_1^{1-2} } \delta ^{(1)}(r) = -a_1\delta ^{(1)}(r),
\end{equation}
where recall $(-1)!!=1$. The $d\to1$ limit of our generalized pseudopotential [Eq. (\ref{eq:generalPseudo_k0}) from the $k\to0$ limit of Eq. (\ref{pseudoPotential})] agrees with the above result without the need to drop the regularizing operator, as we now show. Indeed, operating on a wave function $\Psi$ with our generalized pseudopotential (Eq. \ref{eq:generalPseudo_k0}), the generalized Leibniz product rule (Eq.~\ref{eq:general_Leibniz}) gives
\begin{equation}
\begin{split}
&_{C_0}V_{k\to0}^{(d)}\Psi =\frac{\Omega(d)\bar{a}^{d-2}}{\Gamma(d-2)} \, \delta^{(d)}(r)\,{\; _{C_0}D}_r^{d-2}r^{d-2}\Psi\\
&=
\frac{\Omega(d)\bar{a}^{d-2}}{\Gamma(d-2)} \, \delta^{(d)}(r)\,
\sum_{k=0}^\infty \binom{d-2}{k}(D^{d-2-k}r^{d-2})D^k \Psi
\end{split}
\end{equation}
where the second sum from the Leibniz rule (Eq.~\ref{eq:general_Leibniz}) does not appear above because the upper bound of the sum is $m - 1 = \lceil d - 2\rceil -1= -2<0$. Next, applying the Riemann-Liouville fractional derivative of powers (Eq.~\ref{RLpower}) gives
\begin{equation}
\begin{split}
&_{C_0}V_{k\to0}^{(d)}\Psi \\
&=\frac{\Omega(d)\bar{a}^{d-2}}{\Gamma(d-2)} \, \delta^{(d)}(r)\,
\sum_{k=0}^\infty \frac{\Gamma(d-1)}{\Gamma(k+1)\Gamma(d-k-1)}\frac{\Gamma(d-1)}{\Gamma(k+1)}r^kD^k \Psi\\
&=\Omega(d)\bar{a}^{d-2}\, \delta^{(d)}(r)\,
\sum_{k=0}^\infty \frac{(d-2)\Gamma(d-1)}{\Gamma^2(k+1)\Gamma(d-k-1)}r^kD^k \Psi.
\label{dto1check}
\end{split}
\end{equation}
When $r \to 0$, the only non-zero term in Eq. \ref{dto1check} is $k = 0$. Hence,
\begin{equation}
\begin{split}
_{C_0}V_{k\to0}^{(d)}\Psi& =\Omega(d)\bar{a}^{d-2}\, \delta^{(d)}(r)
\frac{(d-2)\Gamma(d-1)}{\Gamma^2(1)\Gamma(d-1)}D^0 \Psi\\
&=\Omega(d)\bar{a}^{d-2}\, \delta^{(d)}(r)\,
(d-2)\Psi.
\label{}
\end{split}
\end{equation} 
Thus, when $d \to 1$, we obtain
\begin{equation}
_{C_0}V_{k\to0}^{(1)}\Psi =-\frac{2}{\bar{a}_1} \, \delta^{(1)}(r)\Psi.
\label{dto1}
\end{equation} 
Converting our scattering length $\bar{a}_1$ to the W\'odkiewicz constant $a_1$ ($\frac{1}{\bar{a}_1}=\frac{a_1}{2}$)  by Eq. (\ref{aarelationship}), we verify that our pseudopotentials agree for one dimension:
\begin{equation}
_{C_0}V_{k\to0}^{(1)} =-a_1 \delta^{(1)}(r).
\label{dto1}
\end{equation}

\section{Derivation of the $s$-wave $d$-dimensional isotropic harmonic oscillator wave function and energy}\label{appendix:HO}
Here we derive the solution of the $d$-dimensional harmonic oscillator Schr\"odinger equation. We obtain energies and the wave function in terms of the Laguerre polynomial. These non-interacting results are needed as a basis for the trapped energy for two atoms interacting via the generalized pseudopotential,  derived in Appendix \ref{appendixD}. We begin with the $d$-dimensional harmonic oscillator Schr\"odinger equation in oscillator units: 
\begin{equation}
\left[-\frac{1}{2}\left(\frac{d^2}{dr^2} + \frac{d-1}{r}\frac{d}{dr}\right) + \frac{1}{2}r^2 \right]\phi_n^{(d)}(r)=E_n^{(d)}\phi_n^{(d)}.
\end{equation}
Transforming this differential equation into the Laguerre form requires two steps. First we transform the wave function as
\begin{equation}\label{eq:tempTransform}
\phi_n^{(d)} = e^{\frac{-r^2}{2}} \Phi_n^{(d)},
\end{equation}
which leads to
 \begin{equation}
-\frac{1}{2}\left(\frac{d^2}{dr^2} + \frac{d-1-2r^2}{r}\frac{d}{dr} - d \right)\Phi_n^{(d)}(r)=E_n^{(d)}\Phi_n^{(d)}.
\end{equation}
Second we let $u=r^2$ to obtain
\begin{equation}
\left[u\frac{d^2}{du^2} + \left(\frac{d}{2} - u \right)\frac{d}{du} + \left(\frac{E_n^{(d)}}{2}-\frac{d}{4} \right)\right]\Phi_n^{(d)}(u)=0,
\end{equation}
and in order for the above equation to match the associated Laguerre differential equation: 
\begin{equation}
\left[u\frac{d^2}{du^2} + \left(\alpha -u +1 \right)\frac{d}{du} + n \right]L_n^{(\alpha)}(u)=0,
\end{equation}
we require $\alpha = \frac{d-2}{2}$ and the following relationship to hold
\begin{equation}\label{eq:EnHO}
{E_n^{(d)}} = 2n+\frac{d}{2}.
\end{equation}
Using this value for $\alpha$ and Eq. (\ref{eq:tempTransform}), we can write the $d$-dimensional harmonic oscillator wave function in terms of the Laguerre polynomial:
\begin{equation}
\phi_n^{(d)}(r) = A e^{-r^2/2} {L_n}^{\frac{d-2}{2}}(r^2),
\end{equation}
where $A$ is a normalization constant. We use the normalization condition $\int | \phi_n^{(d)} |^2 d\vec{r}^{d} = 1$ to find $A$, which leads to the following integral to solve
\begin{equation}\label{eq:tempAho}
A^2 \Omega(d) \int_0^{\infty}e^{-r^2} \left( {L_n}^{\frac{d-2}{2}}(r^2) \right)^2 r^{d-1}dr = 1.
\end{equation}
We can solve Eq. (\ref{eq:tempAho}) using the similar known relationship for Laguerre polynomials \cite{arfken}:
\begin{equation}
\int_0^{\infty}e^{-x} x^\lambda \left( {L_n}^\lambda(x) \right)^2 dx = \frac{\Gamma\left(n+\lambda+1\right)}{\Gamma\left(n+1\right)}.
\end{equation}
If we let $x=r^2$ and $\lambda=\frac{d-2}{2}$, then
\begin{equation}
\int_0^{\infty}e^{-r^2} \left( {L_n}^{\frac{d-2}{2}}(r^2) \right)^2 r^{d-1}dr = \frac{\Gamma\left(n+d/2\right)}{\Gamma\left(n+1\right)},
\end{equation}
and substituting this for the integral in Eq. (\ref{eq:tempAho}), we find the normalization constant is
\begin{equation}
A= \sqrt{ \frac{2 \Gamma\left(n+1\right)}{\Omega(d) \Gamma\left(n+d/2\right)} },
\end{equation}
and finally the $d$-dimensional harmonic oscillator wave function is
\begin{equation}\label{eq:HO_wf}
\phi_n^{(d)}(r) = \sqrt{ \frac{2 \Gamma\left(n+1\right)}{\Omega(d) \Gamma\left(n+d/2\right)} } e^{-r^2/2} {L_n}^{\frac{d-2}{2}}(r^2).
\end{equation}

Because of the involvement of $\phi_n^{(d)}(0)$ in Eq. (\ref{appC1}) from the derivation of the solution of two trapped atoms interacting via the generalized pseudopotential, it proves useful to write the normalization constant in Eq. (\ref{eq:HO_wf}) in terms of ${L}_n^{\frac{d-2}{2}} (0)$. We do this by noting that
\begin{equation}
 {L_n}^{\frac{d-2}{2}}(0)  = \frac{\Gamma\left(n+d/2\right)}{\Gamma{\left(d/2\right)}\Gamma\left(n+1\right)},
\end{equation}
which can be used in Eq. (\ref{eq:HO_wf}) to obtain
\begin{align}\label{eq:LaguerreWF}
\phi_n^{(d)}(r) & = \sqrt{\frac{2}{\Omega(d)\Gamma(d/2)}} \left( {L}_n^{\frac{d-2}{2}} (0) \right)^{-1/2} e^{-r^2/2} {L_n}^{\frac{d-2}{2}}(r^2).
\end{align}

\section{Details of the construction of the implicit energy equation $E^{(d)}$ for two cold atoms ($k\to0$) in a harmonic trap with pseudopotential interaction in non-integer dimension (Eq. (\ref{eq:energyEquation}))}\label{appendixD}
In this appendix we generalize the Busch derivation in \cite{busch} to arbitrary dimension, including non-integer. We start at Eq. (\ref{eq:startBusch}) and rearrange the terms to obtain:
\begin{equation}
\begin{split}
&\sum_{n=0}^\infty c_n \bigl({E_n^{(d)}}-E^{(d)}\bigr) \phi_n^{(d)}\\
& +\frac{\Omega(d)\bar{a}^{d-2}}{\Gamma(d-2)}\delta^{(d)}(r)\left[{_{C_0}D}_{r}^{d-2} r^{d-2}\sum_{n=0}^\infty c_n\phi_n^{(d)}\right]_{r\rightarrow 0}=0,
\end{split}
\end{equation}
where ${_{C_0}D}_{r}^{d-2}$ is the Caputo fractional derivative and we are using the $k\to0$ generalized pseudopotential (Eq.\ref{eq:generalPseudo_k0}). Projecting both sides onto $\phi_n^{(d)}(\vec{r})$, we have
\begin{equation}
\begin{split}
&c_n({E_n^{(d)}}-E^{(d)})+\frac{\Omega(d)\bar{a}^{d-2}}{\Gamma(d-2)} 
\phi_n^{(d)}(0)\\
& \times\bigl[ {_{C_0}D}_{r}^{d-2}r^{d-2}\sum_{n=0}^\infty c_n\phi_n^{(d)}\bigr]_{r\rightarrow 0}=0,
\end{split}
\end{equation}
and $c_n=A\dfrac{\phi_n^{(d)}(0)}{E^{(d)}_n-E^{(d)}}$, yielding:
\begin{equation}
\begin{split}
&A\phi_n^{(d)}(0)+ \frac{\Omega(d)\bar{a}^{d-2}}{\Gamma(d-2)} \phi_n^{(d)}(0)\\
&\times \bigl[{_{C_0}D}_{r}^{d-2}r^{d-2}\sum_{n=0}^\infty \frac{A\phi_n^{(d)}(0)\phi_n^{(d)}(\vec{r})}{{E_n^{(d)}}-E^{(d)}}\bigr]_{r\rightarrow 0}=0,
\end{split}
\end{equation}
or, dividing by $A\phi_n^{(d)}(0)$,
\begin{equation}
1+ \frac{\Omega(d)\bar{a}^{d-2}}{\Gamma(d-2)} \left[{_{C_0}D}_{r}^{d-2}r^{d-2}\sum_{n=0}^\infty \frac{\phi_n^{(d)}(0)\phi_n^{(d)}(\vec{r})}{{E_n^{(d)}}-E^{(d)}}\right]_{r\rightarrow 0}=0,
\end{equation}
which results in
\begin{equation}
\label{appC1}
\frac{\Omega(d)}{\Gamma(d-2)} \left[ _{C_0}D_r^{d-2}  r^{d-2} \sum_{n=0}^\infty \frac{\phi_n^{(d)}(0)\phi_n^{(d)}(\vec{r})}{{E_n^{(d)}}-E^{(d)}} \right] _{r\rightarrow 0} = -\frac{1}{\bar{a}^{d-2}}.
\end{equation}
In order to simplify Eq. (\ref{appC1}), we insert the representation of the harmonic oscillator $s$-wave functions for arbitrary dimension (Eq. \ref{eq:LaguerreWF}):
\begin{align}
\phi_n^{(d)}(r) & = \sqrt{\frac{2}{\Omega(d)\Gamma(d/2)}} \left( {L}_n^{\frac{d-2}{2}} (0) \right)^{-1/2} e^{-r^2/2} {L_n}^{\frac{d-2}{2}}(r^2),
\end{align}
where ${L}_n^{\frac{d-2}{2}}(r^2)$ is the associated Laguerre polynomial, yielding:
\begin{equation}
\begin{split}
&\frac{2}{\Gamma(d-2)\Gamma(\frac{d}{2})}
\bigl[ _{C_0}D_r^{d-2}
	 r^{d-2} 
		\sum_{n=0}^\infty
		\frac
			{e^{-r^2/2}	{L}_n^{\frac{d-2}{2}}(r^2)}
			{{E_n^{(d)}} - E^{(d)}}
\bigr]_{r\rightarrow 0}\\
&= -\frac{1}{\bar{a}^{d-2}}.
\end{split}
\end{equation}
Recalling the unperturbed harmonic oscillator energies (Eq.~\ref{eq:EnHO})), ${E_n^{(d)}} = 2n+\frac{d}{2}$, we introduce the variable 
\begin{equation}\label{eq:nu_E}
\nu = \frac{E^{(d)}}{2} - \frac{d}{4}
\end{equation} 
as the non-integer equivalent of quantum number $n$ and obtain:
\begin{equation}
\label{appC3}
\begin{split}
&\frac{1}{\Gamma(d-2)\Gamma(\frac{d}{2})}
\bigl[ _{C_0}D_r^{d-2}
	 r^{d-2} e^{-r^2/2}
		\sum_{n=0}^\infty
		\frac
			{{L}_n^{\frac{d-2}{2}}(r^2)}
			{n-\nu}
\bigr]_{r\rightarrow 0}\\
&= -\frac{1}{\bar{a}^{d-2}}.
\end{split}
\end{equation}

Using the identity 
\begin{equation}
\frac{1}{n-\nu} = 
\int_0^\infty
\frac
	{dy}
	{(1+y)^2}
\left(\frac
	{y}
	{1+y}\right) 
^{n-\nu-1} 
 \quad n-\nu > 0,
\end{equation}
the summation becomes
\begin{equation}
\label{appC2}
\sum_{n=0}^\infty
\frac
	{{L}_n^{\frac{d-2}{2}}(r^2)}
	{n-\nu}
=
\sum_{n=0}^\infty
\int_0^\infty
\frac
	{dy}
	{(1+y)^2}
\left(
	\frac{y}{1+y}
\right)^{n-\nu-1}
{L}_n^{	\frac{d-2}{2}	}(r^2).
\end{equation}
We now use the generating functions of the Laguerre Polynomials~\cite{arfken}, 
\begin{equation}
\sum_{n=0}^\infty 
{L}_n^k(x)z^n
=
(1-z)^{-(k+1)}
e^{-xz/(1-z)},
\end{equation}
letting $k=\frac{d-2}{2}$, $x=r^2$, and $z=\frac{y}{1+y}$, to rewrite Eq. \ref{appC2}: 
\begin{equation}
\sum_{n=0}^\infty
\frac
	{{L}_n^{\frac{d-2}{2}}(r^2)}
	{n-\nu}
=
\int_0^\infty
\frac
	{dy}
	{(1+y)^2}
\left(
	\frac
	{y}
	{1+y}
\right)^{-\nu-1}
e^{-r^2y}
(1+y)^{\frac{d}{2}}
\end{equation}
which simplifies to 
\begin{equation}
\sum_{n=0}^\infty
\frac
	{{L}_n^{\frac{d-2}{2}}(r^2)}
	{n-\nu}
=
\int_0^\infty
e^{-r^2y}
y^{-\nu-1}
(1+y)^{\frac{d}{2}+\nu-1}dy.
\end{equation}
Using the integral representation of the confluent hypergeometric function~\cite{handbookMathFuncs}
\begin{equation}
\Gamma(a)U(a,b,z)
=
\int_0^\infty
e^{-zt}
t^{a-1}
(1+t)^{b-a-1}dt,
\end{equation}
where $a=-\nu$, $b=\frac{d}{2}$, $z=r^2$, and $t=y$, Eq. (\ref{appC2}) becomes
\begin{equation}\label{eq:Lag_U}
\sum_{n=0}^\infty
\frac
	{{L}_n^{\frac{d-2}{2}}(r^2)}
	{n-\nu}
=
\Gamma(-\nu)U(-\nu,\frac{d}{2},r^2).
\end{equation}
We further examine the behavior of Eq. (\ref{eq:Lag_U}) as $r \to 0$ by using the identity for $U$ in terms of Kummer's function, $M$ (13.1.2 and 13.1.3 from \cite{handbookMathFuncs}) to get:
\begin{equation}
\begin{split}
&\Gamma(-\nu)U(-\nu,\frac{d}{2},r^2)\\
 & = \frac{\pi\Gamma(-\nu)}{\sin(\frac{d}{2}\pi)}
\bigl(
\frac{M(-\nu, \frac{d}{2}, r^2)}{\Gamma(1- \frac{d}{2}-\nu)\Gamma( \frac{d}{2})}
 - 
r^{2-d}\frac{M(1-\nu- \frac{d}{2}, 2- \frac{d}{2}, r^2)}{\Gamma(-\nu)\Gamma(2- \frac{d}{2})}
 \bigr)\\
 & = \pi\csc\bigl(\frac{d\pi}{2}\bigr)
\bigl(
\frac{\Gamma(-\nu)}{\Gamma(1- \frac{d}{2}-\nu)\Gamma( \frac{d}{2})}
 - 
\frac{r^{2-d}}{\Gamma(2- \frac{d}{2})}\sum_{i = 0}^{\lfloor \frac{d}{2}-1\rfloor}\frac{(1-\nu- \frac{d}{2})_{(i)}}{(2- \frac{d}{2})_{(i)}}r^{2i}
 + O(r) \bigr)
\end{split}
\end{equation}

where $(x)_{(i)}$ is the Pochhammer symbol. Note the Kummer's series above are truncated because they do not contribute following differentiation and the $r\to 0$ limit is applied in Eq. (\ref{appC3}). From here, the $[\cdot]_{r\to 0}$ limit in Eq. (\ref{appC3}) becomes
\begin{equation}
\begin{split}
&\bigl[ _{C_0}D_r^{d-2}
	 r^{d-2} e^{-r^2/2}
		\sum_{n=0}^\infty
		\frac
			{{L}_n^{\frac{d-2}{2}}(r^2)}
			{n-\nu}
\bigr]_{r\rightarrow 0} \\
&= \bigl[D_r^{d-2}
	 r^{d-2} e^{-r^2/2}
		\Gamma(-\nu)U(-\nu,\frac{d}{2},r^2)
\bigr]_{r\rightarrow 0} \\
&= \pi	\csc\bigl(\frac{d\pi}{2}\bigr)
\Bigl[D_r^{d-2}
	e^{-r^2/2}\\
&\bigl(
\frac{\Gamma(-\nu)r^{d-2} }{\Gamma(1- \frac{d}{2}-\nu)\Gamma( \frac{d}{2})}
 - 
\frac{1}{\Gamma(2- \frac{d}{2})} + O(r)
 \bigr)
\Bigr]_{r\rightarrow 0}. \label{eq:D_of_gaussian}
\end{split}
\end{equation}

When $d$ is non-integer, derivatives such as $D_r^{d-2}e^{-r^2/2}r^{d-2}$ above do not follow the usual rules of differentiation. To derive the Caputo fractional derivative, $D_r^{\eta}(e^{-r^2/2}r^{\eta})$, we use the Leibniz product rule (Eq.~\ref{eq:general_Leibniz}) for Caputo derivatives~\cite{ishteva_thesis}:
\begin{equation}
\begin{split}
&D^\eta(f,g)\\
& = \sum_{k=0}^\infty\binom{\eta}{k}(D^{\eta-k}f)g^{(k)} - \sum_{k=0}^{m-1}\frac{r^{k-\eta}}{\Gamma(k+1-\eta)}(fg)^{(k)}(0),
\end{split}
\end{equation}
where $\eta$ may be non-integer and $m = \lceil \eta \rceil$. For notation, we assume $D^*$ derivatives are with respect to $r$ unless indicated otherwise. Applying this product rule gives
\begin{equation}\label{eq:Dgaussian_full}
\begin{split}
D^{\eta}\bigl(e^{-\frac{r^2}{2}}r^{\eta}\bigr)
& = 
 \sum_{k=0}^\infty \bigl(\begin{matrix}\eta\\k\end{matrix}\bigr) 
 	(D^{\eta-k}r^{\eta})D^k\bigl(e^{-\frac{r^2}{2}}\bigr)\\
& - 
 \sum_{k=0}^{m-1}
 	\frac{r^{k-\eta}}{\Gamma(k+1-\eta)}
	D^{k}\bigl(e^{-\frac{r^2}{2}}r^{\eta}\bigr)
\end{split}
\end{equation}

We are interested in this derivative as $r\to0$ for Eq. (\ref{eq:D_of_gaussian}) and ultimately Eq. (\ref{appC3}). Note that the second summation in Eq. (\ref{eq:Dgaussian_full}) involves the integer-order derivatives $D^{k}\left(e^{-\frac{r^2}{2}}r^{\eta}\right)$ of $e^{-\frac{r^2}{2}}$ multiplied with only positive powers of $r$ for any $k \leq m-1 < \eta$. Hence, when evaluated at $r=0$, the terms in this summation are 0, and when $r\to0$, Eq. (\ref{eq:Dgaussian_full}) becomes
\begin{equation}
\begin{split}
&\bigl[ D^{\eta}\bigl(e^{-\frac{r^2}{2}}r^{\eta}\bigr)\bigr]_{r\to0}
= \sum_{k=0}^\infty \bigl(\begin{matrix}\eta\\k\end{matrix}\bigr) 
 	(D^{\eta-k}r^{\eta})D^k\bigl(e^{-\frac{r^2}{2}}\bigr) \\
&={{\eta}\choose{0}}
		(D^\eta r^\eta) e^{-\frac{r^2}{2}} + 
	\sum_{k=1}^\infty \bigl(\begin{matrix}\eta\\k\end{matrix}\bigr) 
		\bigl(D^{\eta-k}r^{\eta}\bigr)\bigl(D^ke^{-\frac{r^2}{2}}\bigr) \\
&=\Gamma(\eta+1) e^{-\frac{r^2}{2}} + 
	\sum_{k=1}^\infty 
	\frac{r^k}{\Gamma(k-\eta+1)}
		\bigl(D^ke^{-\frac{r^2}{2}}\bigr) \label{derivGaussInt}
\end{split}
\end{equation}



We will now show that the summation above (\ref{derivGaussInt}) is 0 as $r \to 0$. The derivative in parentheses in Eq. (\ref{derivGaussInt}) is 
\begin{equation}
\begin{split}
&D^ke^{-\frac{r^2}{2}}
 = 2^{\frac{k}{2}}e^{-\frac{r^2}{2}}(-1)^k U\bigl(-\frac{k}{2}, \frac{1}{2}, \frac{r^2}{2}\bigr)\\
& = 2^{\frac{k}{2}}e^{-\frac{r^2}{2}}(-1)^k \sqrt\pi \bigl(\frac{M\bigl(-\frac{k}{2}, \frac{1}{2}, \frac{r^2}{2}\bigr)}{\Gamma\bigl(\frac{1-k}{2}\bigr)}\\
& - \frac{r\sqrt2 M\bigl(\frac{1-k}{2}, \frac{1}{2}, \frac{r^2}{2}\bigr)}{\Gamma\bigl(\frac{-k}{2}\bigr)} \bigr)
\end{split}
\end{equation}
Since the Kummer's function $M(\cdot, \cdot, \frac{r^2}{2})$ contains non-negative powers of $r$, multiplying this derivative with $r^k$ results in the summation of (\ref{derivGaussInt}) going to 0 as $r \to 0$. Thus, Eq. (\ref{derivGaussInt}) becomes
\begin{equation}
\begin{split}
&\bigl[D^{\eta}\bigl(e^{-\frac{r^2}{2}}r^{\eta}\bigr)\bigr]_{r\to0}\\
&=\bigl[\Gamma(\eta+1) e^{-\frac{r^2}{2}}\\
&+	\cancelto{0}{\sum_{k=1}^\infty 
	\frac{r^k}{\Gamma(k-\eta+1)}
		\bigl(D^ke^{-\frac{r^2}{2}} \bigr)} \bigr]_{r\to0}\\
& = \Gamma(\eta+1).
\end{split}
\end{equation}
and

\begin{equation}\label{eq:Dgauss_Gamma_dminus1}
\left[D^{d-2}\left(e^{-\frac{r^2}{2}}r^{d-2}\right)\right]_{r\to0}=
		 \Gamma(d-1).
\end{equation}
Similarly, using the Leibniz product rule and letting $g(r) = 1$, we can show that the second term in Eq. (\ref{eq:D_of_gaussian}) is 0 because 
\begin{equation}\label{eq:Dgauss_0}
\left[D^{d-2}\left(e^{-\frac{r^2}{2}}\right)\right]_{r\to0}=0.
\end{equation}
Therefore, from Eqs.(\ref{eq:Dgauss_Gamma_dminus1}) and (\ref{eq:Dgauss_0}) we conclude that Eq. (\ref{eq:D_of_gaussian}) becomes 
\begin{equation}
\begin{split}
&\bigl[ _{C_0}D_r^{d-2}
	 r^{d-2} e^{-r^2/2}
		\sum_{n=0}^\infty
		\frac
			{{L}_n^{\frac{d-2}{2}}(r^2)}
			{n-\nu}
\bigr]_{r\rightarrow 0}\\ 
&= \pi	\csc\bigl(\frac{d\pi}{2}\bigr)
\frac{\Gamma(-\nu)\Gamma(d-1) }{\Gamma(1- \frac{d}{2}-\nu)\Gamma( \frac{d}{2})},
\label{limitSimplified}
\end{split}
\end{equation}
and Eq. (\ref{appC3}) becomes
\begin{equation}
\begin{split}
&\frac{1}{\bar{a}^{d-2}}  = -\frac{1}{\Gamma(d-2)\Gamma(\frac{d}{2})}\\
& \times\pi\csc\left(\frac{d\pi}{2}\right)
\frac{\Gamma(-\nu)\Gamma(d-1) }{\Gamma(1- \frac{d}{2}-\nu)\Gamma( \frac{d}{2})}\\
& = -
 	\csc\left(\frac{d\pi}{2}\right) \frac{\pi(d-2) }{\Gamma^2( \frac{d}{2})}
\frac{\Gamma(-\nu)}{\Gamma(1- \frac{d}{2}-\nu)}.
\end{split}
\end{equation}
Finally, using the relation for $\nu$ (Eq.~\ref{eq:nu_E}), 
\begin{equation}\label{eq:D_2atomEnergy}
\frac{-\sin\left(\frac{d\pi}{2}\right) }{\bar{a}^{d-2}}  =
 	\frac{\pi(d-2) }{\Gamma^2( \frac{d}{2})}
\frac{\Gamma(-\frac{E^{(d)}}{2}+\frac{d}{4})}{\Gamma(-\frac{E^{(d)}}{2} + \frac{4-d}{4})}.
\end{equation}
In Section \ref{sec:5_two_cold} we convert our $\bar{a}_{d}$ to the relative motion units $a_o$ used in Ref.~\cite{busch} via Eq. (\ref{eq:La0}) to make comparison easier.

\section{Weak interaction perturbation solution for the $d$-dimension energy equation for two cold atoms in a harmonic trap}\label{appendixE}
Starting with Eq. (\ref{eq:energyEquation_temp}) (derived in Eq. \ref{eq:D_2atomEnergy}), we have 
\begin{equation}
\begin{split}
&\frac
	{\pi(d-2)}
	{\Gamma^2(\frac{d}{2})}
\dfrac
	{\Gamma(-\frac{E^{(d)}}{2}+\frac{d}{4})}
	{\Gamma(-\frac{E^{(d)}}{2}+\frac{4-d}{4})}\\
& = 
\frac
	{-\sin\left(\frac{d\pi}{2}\right)}
	{\bar{a}^{d-2}_{d}}.
\end{split}
\end{equation}
For notational simplicity, below we drop the $d$ subscript in the dimension dependent radius, $\bar{a}_d$. 

First, we start with the substitution $\epsilon=-\nu = -\frac{E}{2}+\frac{d}{4}$, move the four to the RHS, and take the reciprocal to yield
\begin{equation}
\begin{split}
&\frac
	{\Gamma
	\left(
		\epsilon+1-\frac{d}{2}
	\right)}
	{\Gamma(\epsilon)}\\
&=-\csc\left(\frac{d\pi}{2}\right)
\frac{\pi(d-2)\bar{a}^{d-2}}{\Gamma^2(\frac{d}{2})}.
\end{split}
\end{equation}
In order to expand about $\bar{a}\to 0$, we first observe that the LHS also goes to zero when the gamma function in the denominator approaches poles, when $\epsilon=-n$ for $n\in \mathbb{Z}$. Therefore, we will expand $\epsilon$ about $\epsilon = -n$ to obtain $\epsilon=-n+\bar{a}^{d-2}\epsilon_1$. In effect, we are trying to evaluate
\begin{equation}
\left[
\frac
	{\Gamma
	\left(
	\epsilon+1-\frac{d}{2}
	\right)}
	{\Gamma(\epsilon)}
\right]
_{\epsilon\rightarrow -n}.
\end{equation}
Using the multiplicative identity, 

\begin{equation}
\begin{split}
&\bigl[
\frac
	{\Gamma
	\bigl(
		\epsilon+1-\frac{d}{2}
	\bigr)}
	{\Gamma(\epsilon)}
\bigr]
_{\epsilon\rightarrow -n}
=
\bigl[
\frac
	{
	\Gamma
	\bigl(
		\epsilon+1-\frac{d}{2}
	\bigr)
	(\epsilon+n)
	}
	{
	\Gamma(\epsilon)
	(\epsilon+n)
	}
\bigr]
_{\epsilon\rightarrow -n}\\
&=
\frac
	{\Gamma
	\bigl(
		-n+1-\frac{d}{2}
	\bigr)
	(\epsilon+n)
	}
	{\bigl[
	\Gamma(\epsilon)
	(\epsilon+n)
	\bigr]
	_{\epsilon \rightarrow -n}
	}
.
\end{split}
\end{equation}
Using $Res(f,z_0)=\lim_{z\to z_0}(z-z_0)f(z)$ and that $\Gamma(-n)$ is a pole with $Res(\Gamma(\epsilon),-n)=(-1)^n/\Gamma(n+1)$, we have


\begin{equation}
\begin{split}
&\bigl[
\frac
	{\Gamma
	\bigl(
		\epsilon+1-\frac{d}{2}
	\bigr)}
	{\Gamma(\epsilon)}
\bigr]
_{\epsilon\rightarrow -n}\\
&=
\frac
	{
	\Gamma \bigl(-n+1-\frac{d}{2}\bigr)
	(\epsilon+n)
	}
	{
	(-1)^n / \Gamma(n+1)
	}
\end{split}
\end{equation}
which simplifies to 
\begin{equation}
\begin{split}
&\bigl[
\frac
	{\Gamma
	\bigl(
		\epsilon+1-\frac{d}{2}
	\bigr)}
	{\Gamma(\epsilon)}
\bigr]
_{\epsilon\rightarrow -n}\\
&=
(-1)^n
\Gamma(n+1)
\Gamma\bigl(-n+1-\frac{d}{2}\bigr)
(\epsilon+n).
\end{split}
\end{equation}
Exploiting the gamma function identity 
\begin{equation}
(-1)^n\Gamma(z-n)=
\frac
	{\Gamma(z)\Gamma(1-z)}
	{\Gamma(n+1-z)}
\end{equation}
where in this instance, $z=1-\frac{d}{2}$, our approximation becomes
\begin{equation}
\begin{split}
&\bigl[
\frac
	{\Gamma
	\bigl(
		\epsilon+1-\frac{d}{2}
	\bigr)}
	{\Gamma(\epsilon)}
\bigr]
_{\epsilon\rightarrow -n}\\
&=
\frac
	{
	\Gamma \bigl(n+1\bigr)
	\Gamma \bigl(\frac{d}{2}\bigr)
	\Gamma \bigl(1-\frac{d}{2}\bigr)
	}
	{
	\Gamma \bigl(n+\frac{d}{2}\bigr)
	}
(\epsilon+n),
\end{split}
\end{equation}
Recall that $\epsilon=-n+\bar{a}^{d-2}\epsilon_1$, we substitute $\epsilon+n = \bar{a}^{d-2}\epsilon_1$ to get the approximation
\begin{equation}
\begin{split}
&-\csc\bigl(\frac{d\pi}{2}\bigr)\frac{\pi(d-2)\bar{a}^{d-2}}{\Gamma^2(\frac{d}{2})}
=
\bigl[
\frac
	{\Gamma
	\bigl(
		\epsilon+1-\frac{d}{2}
	\bigr)}
	{\Gamma(\epsilon)}
\bigr]
_{\epsilon\rightarrow -n}\\
&\approx
\frac
	{
	\Gamma \bigl(n+1\bigr)
	\Gamma \bigl(\frac{d}{2}\bigr)
	\Gamma \bigl(1-\frac{d}{2}\bigr)
	}
	{
	\Gamma \bigl(n+\frac{d}{2}\bigr)
	}
\bar{a}^{d-2}\epsilon_1,
\end{split}
\end{equation}
therefore,
\begin{equation}
\epsilon_1
=-\csc\left(\frac{d\pi}{2}\right)
\frac{\pi(d-2)}{\Gamma^3\left(\frac{d}{2}\right)}
\frac
	{
	\Gamma \left(n+\frac{d}{2}\right)
	}
	{
	\Gamma \left(n+1\right)
	\Gamma \left(1-\frac{d}{2}\right)
	}.
\end{equation}
Our final approximation for $\epsilon=-n+\bar{a}^{d-2}\epsilon_1$ becomes
\begin{equation}
\epsilon = -n 
-\bar{a}^{d-2}
\frac{\pi(d-2)}{\sin\left(\frac{d\pi}{2}\right)\Gamma^3\left(\frac{d}{2}\right)}
\frac
	{
	\Gamma \left(n+\frac{d}{2}\right)
	}
	{
	\Gamma \left(n+1\right)
	\Gamma \left(1-\frac{d}{2}\right)
	}.
\end{equation}
and by substituting $\epsilon=-\frac{E}{2}+\frac{d}{4}$ and solving for $E$, our Taylor series expansion for energy $E$ about small $\bar{a}$ is
\begin{equation}
\begin{split}
&E^{(d)}= 2n +\frac{d}{2}
+
\frac{2\pi(d-2)}{\sin\bigl(\frac{d\pi}{2}\bigr)\Gamma^3\bigl(\frac{d}{2}\bigr)}\\
&\frac
	{
	\Gamma \bigl(n+\frac{d}{2}\bigr)
	}
	{
	\Gamma \bigl(n+1\bigr)
	\Gamma \bigl(1-\frac{d}{2}\bigr) 
	} \bar{a}^{d-2}_{d}.
\end{split}
\end{equation}
In Section \ref{sec:5_two_cold} we verify that the $d=1,2,3$ results match Ref.~\cite{busch} by replacing their factorials in the binomial coefficient with gamma functions and converting our $\bar{a}_{d}$ to their units $a_o$ via Eq. (\ref{eq:La0}).


\subsection{Expansion about $d=2$}
The $d\to2$ limit for the energy requires special consideration. Using the scattering length unit conversion (\ref{eq:La0}), our exact energy functional in terms of Busch's $a_o$ is
\begin{equation}
 	\frac{\pi(d-2)}{\Gamma^2( \frac{d}{2})}
\frac{\Gamma(-\frac{E^{(d)}}{2}+\frac{d}{4})}{\Gamma(-\frac{E^{(d)}}{2} + \frac{4-d}{4})} = 
					\frac{-\sin\left(\frac{d\pi}{2}\right)} 
								{ {{a_o}}^{d-2}/2^{d/2} }.
\end{equation} 
Letting $d=2$ in $\Gamma^2( \frac{d}{2})$ and $2^{d/2}$ and rearranging, the above equation becomes
\begin{equation}
 	-\frac{\sin\left(\frac{d\pi}{2}\right)}{d-2} \frac{\Gamma(-\frac{E^{(d)}}{2} + \frac{4-d}{4})}{\Gamma(-\frac{E^{(d)}}{2}+\frac{d}{4})} = 
					\pi \frac{{{a_o}}^{d-2}}{ 2 }.
\end{equation}
Expanding both sides about $d=2$, to first order in $d-2$ we find
\begin{equation}
\begin{split}
&\frac{\pi}{2} + \frac{\pi}{4}(d-2)2\ln{a_o}+\ldots\\
& = \frac{\pi}{2} - \frac{\pi}{4}(d-2) \psi \bigl(\frac{1}{2}-\frac{E^{(2)}}{2} \bigr) + \ldots,
\end{split}
\end{equation}
where $\psi(\cdot)$ is the logarithmic derivative of Euler's $\Gamma$-function. Finally, the first order terms yield the first-order approximation to the $2d$ energy:
\begin{equation}\label{eq:busch2d_derived}
\psi \left(\frac{1}{2} - \frac{E^{(2)}}{2} \right) = \ln \left( \frac{1}{{{a_o}}^2} \right).
\end{equation}

\section{Typographical errors in W\'odkiewicz Ref.~\cite{wod}}\label{typoAppendix}
\begin{enumerate}
\item In calculating the scatter wave (Eq. (2.8)), the complex parameter $z$ should be evaluated at $z = e^{-i\pi/2k^2}/(4\alpha)$, not $z = e^{i\pi/2k^2}/4\alpha$.

\item Derived from Eq. (2.8), Eq. (2.10) should have no negative sign in the first exponential term:
$$\lim_{r\to \infty} \psi_k^{(+)}(r)=e^{ik\cdot r} + f_k^{(d)}\frac{e^{ikr}}{r^{d/2-1/2}}$$
This correction matches Eqs. (3.2a), (3.5a) and (3.9a). 

\item Eq. (3.1) should read 
$$f_k^{(d)}=-\frac{i^{-n} k^{n-1} (2 n-1)\text{!!} }{k^{2 n-1}+i  2^{n-1} \frac{\pi ^n (2 n-1)\text{!!}\hbar}{  a_{2 n+1} \alpha}}$$
This formula is the result of the derivation described in the paper and also matches with the rest of the paper's Eqs. (3.2b), (3.5b) and (3.9b).

\item The symbol $k$ in Eq. (5.5) should be $l$.
\end{enumerate}

\end{document}